\documentclass[acmsmall,nonacm]{acmart}

\acmDOI{}
\setcopyright{none}
\renewcommand\footnotetextcopyrightpermission[1]{}
\usepackage[T1]{fontenc}

\usepackage[english]{babel}
\usepackage{graphicx}
\usepackage{wrapfig}
\usepackage{booktabs}
\usepackage{xspace}
\usepackage[utf8]{inputenc}
\usepackage{wasysym}
\usepackage[ruled,vlined]{algorithm2e}
\usepackage{pdflscape}
\usepackage{enumitem}
\usepackage{verbatim}
\usepackage{verbatim}
\usepackage{floatflt} 

\usepackage{subcaption}
\usepackage{balance}
\usepackage{float}
\usepackage{xcolor}
\colorlet{RED}{red}
\usepackage{color}
\usepackage{colortbl}
\SetExtraKerning{encoding=*,font=*}{\textemdash={120,120}} 

\newcommand{\todo}[1]{}
\renewcommand{\todo}[1]{{\color{red} TODO: {#1}}}

\def \paragraph [#1] {\vspace{3pt}\noindent{\textbf{#1}\quad}}%

\newcommand{\netflix}{Netflix\xspace}

\setcopyright{none}
\begin{document}
\pagestyle{plain}

\makeatletter
\let\@authorsaddresses\@empty
\makeatother

\title{A Global Perspective on the Past, Present, and Future of Video Streaming over Starlink\\
}

\author[]{Liz Izhikevich}
\affiliation[]{\institution{University of California, Los Angeles}
\country{USA}}
\email{lizhikev@ucla.edu}

\author[]{Reese Enghardt}
\affiliation[]{\institution{Netflix}
\country{USA}}

\author[]{Te-Yuan Huang}
\affiliation[]{\institution{Netflix}
\country{USA}}

\author[]{Renata Teixeira}
\affiliation[]{\institution{Netflix}
\country{USA}}



\begin{abstract}

This study presents the first global analysis of on-demand video streaming over Low Earth Orbit (LEO) satellite networks, using data from over one million households across 85 countries. We highlight Starlink's role as a major LEO provider, enhancing connectivity in underserved regions. Our findings reveal that while overall video quality on Starlink matches that of traditional networks, the inherent variability in LEO conditions---such as throughput fluctuations and packet loss---leads to an increase in bitrate switches and rebuffers.
To further improve the quality of experience for the LEO community, we manipulate existing congestion control and adaptive bitrate streaming algorithms using simulation and real A/B tests deployed on over one million households. 
Our results underscore the need for video streaming and congestion control algorithms to adapt to rapidly evolving network landscapes, ensuring high-quality service across diverse and dynamic network types.
\looseness=-1

\end{abstract}

\maketitle
\thispagestyle{empty}

\section{Introduction}
Over the past 10~years, users streaming on-demand video continue to generate the majority of Internet traffic~\cite{San2023}.
However, users are adopting new ways to stream video.
Today, millions of users utilize Low Earth Orbit satellite networks (``LEO'') for Internet connectivity~\cite{spaceX-2mil}.
LEO is used for primary connectivity in remote locations~\cite{remote_labs,under_prvlgd_schools} and as an emergency backup in urban areas~\cite{corporations_star_backup,ind2_star_backup}.

LEO creates a unique environment for video streaming.
LEO's highly-mobile core infrastructure produces sudden changes in latency~\cite{izhikevich2024democratizing}, throughput~\cite{mohan2023multifaceted}, and availability~\cite{ma2023network} that are distinct from traditional networks. 
While video streaming is built to withstand fluctuations in network performance~\cite{yin2015control}, LEO forces video delivery to adapt to more frequent and sudden changes in network conditions than ever before. 

As LEO's popularity continues to grow, it is important to understand how existing video streaming algorithms interact with LEO and where they can be improved.  
However, understanding operational LEO networks presents a challenge due to high barriers to measurement, often requiring researchers to recruit large numbers of satellite Internet customers or deploy specialized hardware~\cite{izhikevich2024democratizing}.
Consequently, video streaming over LEO has remained relatively opaque---studied in small-scale experiments with few users \cite{mohan2023multifaceted,zhao2023realtime,zhao2024low}---and algorithmic advances to improve data delivery over LEO have remained largely theoretical \cite{page2023distributed,cao2023satcp}.

In this work, we provide the first global analysis of on-demand video streaming over LEO.
Through collaboration with a large video streaming service, \netflix,
%
%
we analyze over one million LEO households across 85~countries for over two years.
We show that LEO is a key network for video delivery; as of August~2024, Starlink~\cite{starlink}, the largest LEO provider to date, ranks 29th of 20,000 ISPs that deliver the most seconds of \netflix's streamed video content.
\netflix's perspective finds that Starlink's global presence is unique, responsible for the most unique countries streaming video compared to any other ISP.


We find that overall perceptual video quality for customers streaming \netflix over LEO is surprisingly similar to non-LEO networks in well-served territories, and better than those in non-LEO networks in historically underserved territories (e.g., Africa and small islands).
Nevertheless, Starlink's highly-variable throughput and non-congestive packet loss lead to a marginal increase in bitrate switches and rebuffers, which disproportionately affects under-served territories. 
We manipulate existing congestion control (Section~\ref{sec:tcp}) and adaptive bitrate streaming algorithms (Section~\ref{sec:abr_design}) through simulations and real A/B tests deployed on over one million households, to further improve quality of experience for the global community. 
However, we show that existing design principles are not completely adaptable to LEO's unique nature and present undesirable and inevitable tradeoffs. 

As we look to the future of the Internet, LEO demonstrates the opportunity to provide high quality video streaming to the entire world. 
Yet, our global perspective shows us how video streaming must evolve to simultaneously withstand a diversity of network types (LEO, GEO, Terrestrial, Mobile)
and networks that constantly change their topology and routing.

\section{LEO's Rise in Delivering Video}
\label{sec:leo_rise}

Video streaming over LEO is rapidly growing in popularity and coverage, making it critical to study. 
In this section, we analyze how \netflix streaming over LEO has changed from January 2022--April 2024, and how its growth impacts future LEO research. 
We find that the adoption of Starlink continues to increase, making it the most popular LEO ISP among \netflix's users.
While the most seconds of video streamed over Starlink is by users in the United States, users in Africa and small islands stream the largest \textit{proportion} of video over Starlink compared to non-Starlink networks.
Further, Africa is the fastest growing region for video streaming over Starlink, even though 63\% of the population remained unconnected to the Internet in 2023~\cite{ITU2023}.
If this growth continues at the same rate, Starlink will become a ``top 20'' ISP responsible for the most viewing seconds by April~2025. 


\vspace{3pt}
\noindent
\textbf{Data Collection and Ethics.}
We utilize viewing and performance data traces collected by \netflix, a large video service provider.
The data collected focuses on viewing duration (without looking into what people are viewing), quality of experience metrics (e.g., play delay) collected from the client device, network metrics (e.g., round-trip time) collected from the server's TCP stack, and metadata (e.g., autonomous system, country, device type). 
The traces are collected by \netflix solely to enhance its service. 
Importantly, we ensure that no personally identifiable information is used in our analysis.
We identify Starlink households by their autonomous system (ASN\,14593). 
We describe additional details (e.g., how metrics are specifically measured) in future sections.  


\vspace{3pt}
\noindent
\textbf{Starlink is rapidly growing in popularity.}
Users stream orders magnitude more video over Starlink than any other LEO network.
As of April~2024, the most widely deployed LEO network is Starlink, with near 6K~satellites~\cite{satellite_count} and over 2.6~million users~\cite{nathan_latency_report}.
We limit our study to Starlink, as not enough users stream \netflix over other LEO networks to garner a sufficiently large sample size. 
\looseness=-1

\begin{figure*}[t]
    \centering
    \begin{subfigure}[t]{0.49\textwidth} 
        \vtop{\vskip0pt
        \hbox{\includegraphics[width=\linewidth]{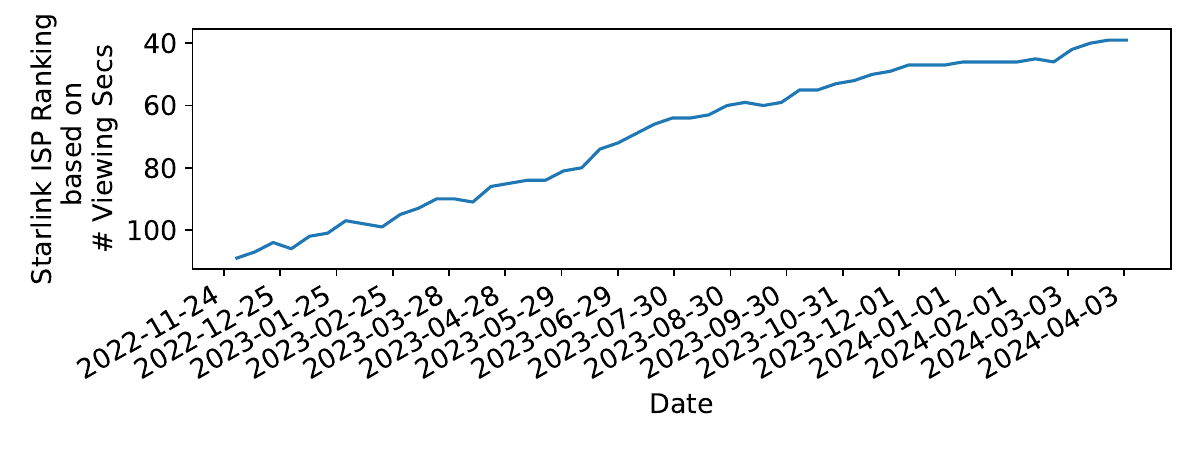}}}
        \caption{\textbf{Starlink is growing in popularity, making it critical to study}---%
        \textnormal{If Starlink continues to grow in popularity at its current rate, it can become a top-20 ISP responsible for the most number of viewing seconds by April~2025.}}
        \label{fig:popularity_overtime}
    \end{subfigure}
    \hfill
    \begin{subfigure}[t]{0.49\textwidth} 
        \vtop{\vskip0pt
        \hbox{\includegraphics[width=\linewidth]{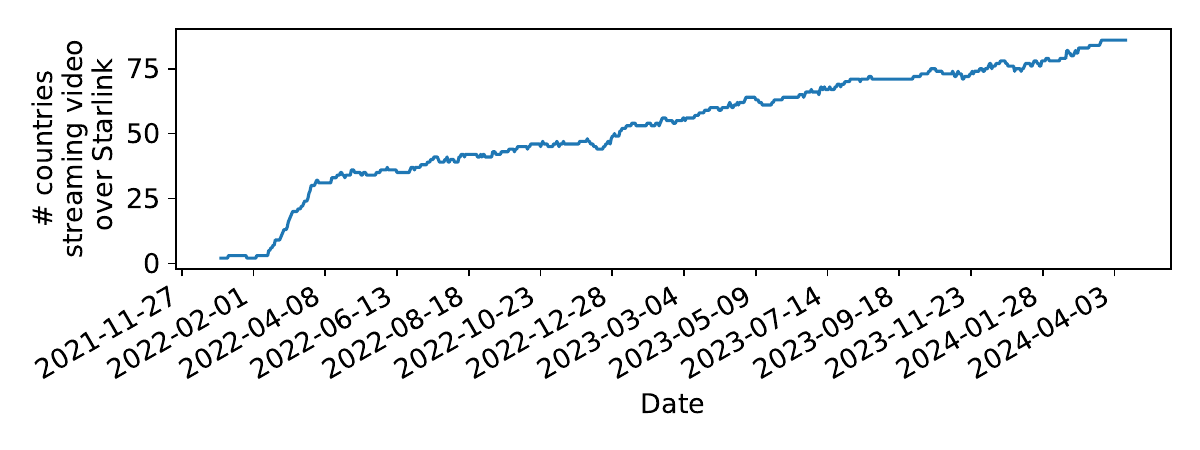}}}
        \caption{\textbf{Starlink is a global ISP}---%
        \textnormal{As of April~2024, customers across over 80 countries stream video over the Starlink network.}}
        \label{fig:countries_stream}
    \end{subfigure}
    \caption{\textbf{Starlink's Growing Popularity and Global Reach}}
    \label{fig:combined_starlink}
\end{figure*}

\begin{figure*}[t]
    \centering
    \begin{subfigure}[t]{0.49\textwidth} 
        \vtop{\vskip0pt 
        \hbox{\includegraphics[width=\linewidth]{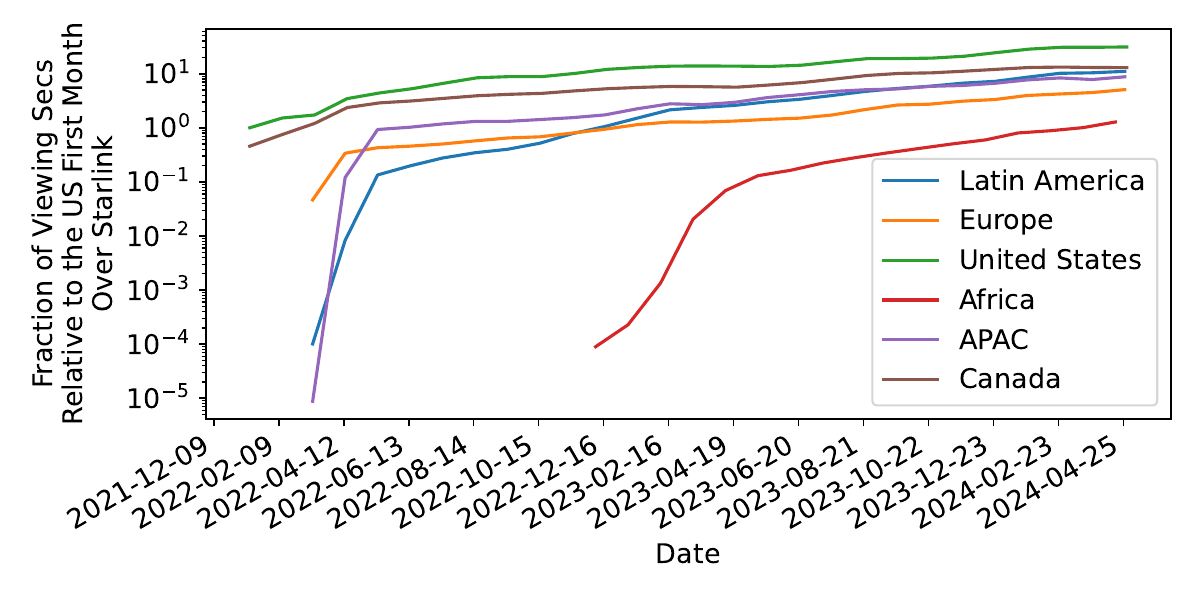}}}
        \caption{\textbf{Regional Popularity of Video Streaming Over Starlink Over Time}---%
        \textnormal{Since Starlink's introduction, users within the United States and Canada dominate video streaming. However, regional trends indicate that streaming in Asia Pacific and Latin America will become more popular than in Canada. The Y-axis shows the fraction of relative growth of viewing and is computed by, for each region, dividing the number of viewing seconds for the denoted month by the number of viewing seconds during the first month that users streamed \netflix over Starlink within the US.}}
        \label{fig:regions_overtime}
    \end{subfigure}
    \hfill
    \begin{subfigure}[t]{0.49\textwidth} 
        \vtop{\vskip0pt 
        \hbox{\includegraphics[width=\linewidth]{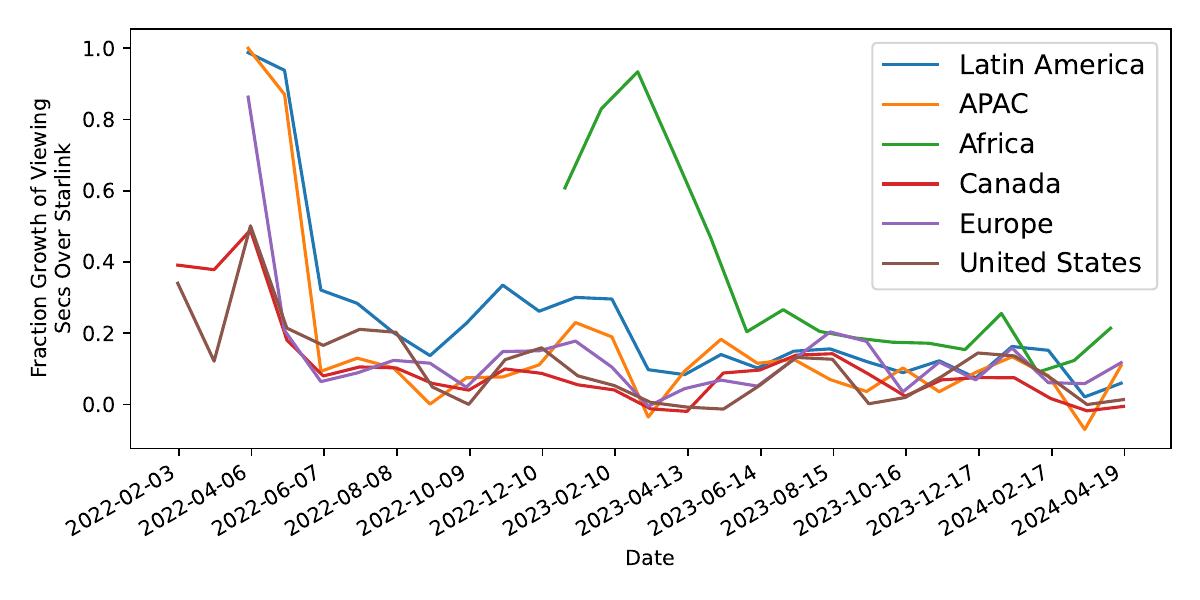}}}
        \caption{\textbf{Relative Video Streaming Growth Across Regions}---%
        \textnormal{Africa is the fastest growing region for video streaming over Starlink, illustrated by the green line remaining towards the top in recent months. The Y-axis shows the fraction of relative growth of viewing relative to the prior month and is computed by, for each region, dividing the number of viewing seconds for the denoted month by the number of viewing seconds during the prior month.}}
        \label{fig:regions_growth}
    \end{subfigure}
    \caption{\textbf{Comparative Analysis of Regional Video Streaming Over Starlink}}
    \label{fig:combined_figure}
\end{figure*}



%
As of August~2024, Starlink is the 29th of over 20,000 ISPs responsible for all global viewing seconds.
In Figure~\ref{fig:popularity_overtime}, we illustrate how Starlink has grown in popularity since 2022 relative to all other ISPs. 
If Starlink continues to grow in popularity at its current rate, it is on track to become a ``top 20'' ISP serving the most \netflix view hours by April~2025. 
Thus, research into operational LEO---beyond the small handful of studies that exist today (Section~\ref{sec:related})---will likely increase in demand and importance.

\vspace{3pt}
\noindent
\textbf{Starlink is a world-wide ISP.}\quad
As Starlink continues to expand its coverage, Starlink customers stream from the largest number of unique countries.\footnote{\netflix geolocates customers by (1) extracting the public IP address they stream from and (2) using multiple commercially available databases to geolocate their IP address.}
In Figure~\ref{fig:countries_stream}, we plot the number of unique countries customers stream from over time. 
As of April~2024, customers from over 85~countries stream video over Starlink. 
Starlink leads in geographic diversity; the second most geographically diverse network
has users from less than half the number countries as Starlink. 
Consequently, obtaining diversity in vantage points will remain critical and challenging for understanding even a single LEO network.

Starlink users within the United States stream the most 
\netflix relative to any other geographic region of Starlink users. 
In April~2024, large mainland territories---the US, Canada, Australia, Mexico, and Brazil---compose the top~5 countries that contribute the most \netflix viewing seconds over Starlink.
Together, they contribute near 90\% of all viewing seconds over Starlink. 
In Figure~\ref{fig:regions_overtime}, we plot the fraction of viewing seconds originating from all geographic regions overtime, relative to the United States in the first month that users streamed \netflix over Starlink. 
As Starlink expands its coverage, different geographic regions quickly begin to increase their video streaming. 
Africa is the newest region to start streaming video, as of December~2022.

Africa is the fastest growing region for video streaming over Starlink, as of April~2024.
In Figure~\ref{fig:regions_growth}, we plot the relative video streaming growth every month compared to the prior month for each region. 
While the majority of regions today increase the number of viewing seconds by roughly 10\%, Africa grows its streaming by over 20\%.
Further, relative to other regions, Africa experiences larger growth (e.g., over 50\% growth) for longer (i.e., 4~months instead of 2 months). 
At this rate, within a year, Africa could soon exceed the number of viewing seconds over Starlink compared to other regions if Starlink continues to deploy across Africa.

\label{sec:rely_dom_country}
Countries in Africa and small islands are most reliant on the Starlink network, relative to local network alternatives. 
Customers on or near the Pitcairn Islands rely on the Starlink most, streaming 30\% of their video over Starlink.
Zambia, Rwanda and Malawi also rely the most on LEO, streaming over 5\% of their video over Starlink. 
In comparison, the United States streams less than 1\% of their video over Starlink. 
Thus, while Islands and African countries only contribute to a small fraction to the overall amount of video streamed over Starlink, maintaining a high quality of experience for these regions is critical as a growing population will be reliant on it.

\section{Quality of Experience}
\label{sec:qoe}

In this section, we analyze how the quality of experience of video streaming over Starlink compares to non-LEO terrestrial networks.
We find that over the years, overall perceptual video quality over Starlink has grown to be the same, and sometimes better, than streaming from alternative networks. 
Yet, at the same time, Starlink customers are 60\% more likely to experience a bitrate switch, and 200\% more likely to experience a rebuffer relative to the most popular Internet service providers. 


\subsection{Methodology}
\label{sub:sec:meth:qoe}
To study quality of experience, we filter for video streaming sessions that are 
\begin{enumerate}
    \item theoretically capable---as determined by \netflix subscription plan and device type---to stream at least at a 720p high definition resolution to ensure low quality is due to the network;
    \item at least 5~minutes long to ensure there is enough data to summarize the quality of a session;
    \item destined towards TVs---as determined by the information the device sends to our servers at the start of a viewing session---which are more likely to be stationary, thereby minimizing customer mobility artifacts; and
    \item streamed during the first week of April~2024, unless otherwise noted. 
\end{enumerate}
Notably, filters 1--3 capture the majority of streaming sessions and filters 1--4 capture millions of streaming sessions for over 1-million Starlink households.
All clients are exposed to the same networking and video algorithms (e.g., congestion control, adaptive bitrate, etc). 

We compare Starlink streaming quality of experience to ``Not Starlink'' networks and the ``Top 10 ISPs'' responsible for the most number of streaming seconds worldwide.
As we will show, Top 10 ISPs provide a marginally better quality of streaming experience compared to all networks and thus act as a ``gold standard.'' 

We quantify \textit{overall} perceptual video quality using a popular full-reference objective video quality assessment algorithm~\cite{saha2023perceptual}: Video Multi-Method Assessment Fusion (VMAF)~\cite{VMAF}. 
The open source algorithm uses machine learning to predict, given a particular video at a particular bitrate and resolution, how humans will perceive its overall quality. 
Critically, VMAF takes into consideration that 
(1) higher video bitrate matters more for certain types of video (action) over others (stationary interviews) and 
(2) increases in bitrate matter more when overall bitrate is low (e.g., 200~Kb/s to 400~Kb/s) as opposed to already being high (e.g., 10~Mb/s  to 10.2~Mb/s). 
The higher the VMAF score, the better the overall perceptual video quality. 
All content available for streaming is annotated with its associated VMAF score at  encoding time. 
VMAF is computed for each encoded video chunk, for all possible representations of the video content. 

Since our goal is to evaluate the network's (in particular, Starlink's) effect on the achieved quality, we report VMAF as a  \textit{Maximum Quality Ratio}.
The Maximum Quality Ratio addresses the fact that some video titles may have a maximum VMAF of 90--due to complexity in encoding--whereas others may have a max VMAF of 100. 
Thus, Maximum Quality Ratio captures whether the video is playing at its best possible quality.
Maximum Quality Ratio is computed by dividing the time-weighted VMAF--- which captures the overall VMAF score across all played chunks (e.g., 2~hour movie)---by the maximum possible VMAF available for the session. 

To further understand what other factors contribute to quality of experience during streaming, we also study throughput and bitrate switches (affecting video quality selection), latency (affecting throughput, play delay, and the reaction time to sudden network changes), and network rebuffers (the ``worst'' video quality). 


\begin{wrapfigure}{t}{0.5\textwidth}
  \vspace{-10pt}
    \includegraphics[scale=0.35]{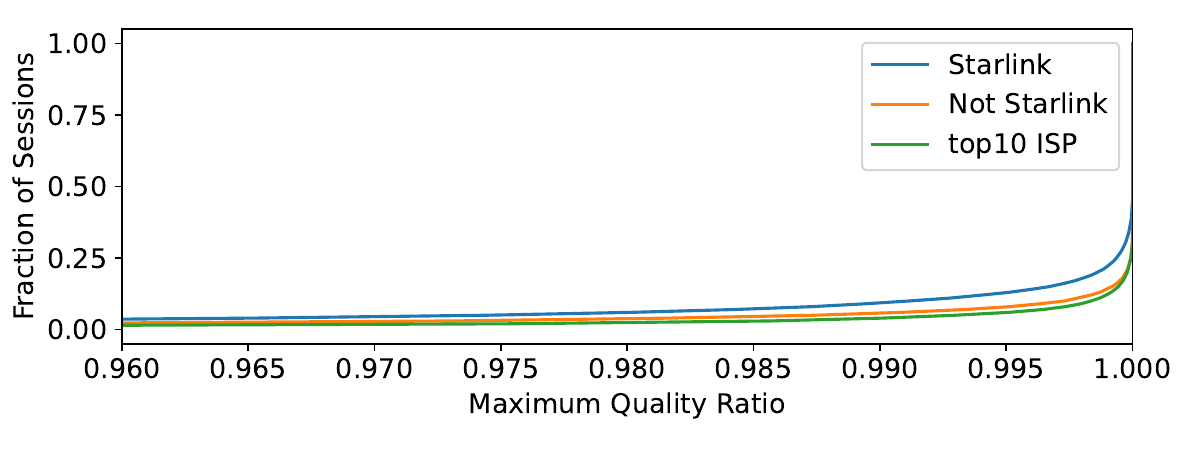}
    \caption{\textbf{Maximum Quality Ratio (Time-weighted VMAF divided by maximum possible VMAF)}---%
    \textnormal{In April~2024, globally, Starlink users experience similar perceptual video quality to users streaming from alternative networks. }}
    \label{fig:MQR_dist}
     \vspace{-10pt}
\end{wrapfigure}

\subsection{Perceptual Video Quality}
\label{sub:sec:vmaf}

Globally, Starlink users often experience similar perceptual video quality compared to users streaming from alternative non-LEO networks. In Figure~\ref{fig:MQR_dist}, we present the distribution of the Maximum Quality Ratio for sessions on Starlink, non-Starlink, and the top 10 ISPs during the first week of April 2024. The x-axis starts at 0.96, indicating that video sessions frequently achieve near-maximum quality for almost their entire duration. Notably, half of the sessions maintain the maximum possible quality throughout their entire duration, with 49\% of sessions for both Starlink and non-Starlink, and 51\% for the top 10 ISPs achieving this benchmark. However, Starlink exhibits a higher percentage of sessions with slightly lower quality; specifically, 10\% of Starlink sessions have a Maximum Quality Ratio below 0.99, compared to only 6\% for non-Starlink ISPs and 4\% for the top 10 ISPs.

\begin{figure*}[t]
    \centering
    \vspace{-10pt}
    \begin{subfigure}[t]{0.49\textwidth}
        \centering
        \includegraphics[scale=0.35]{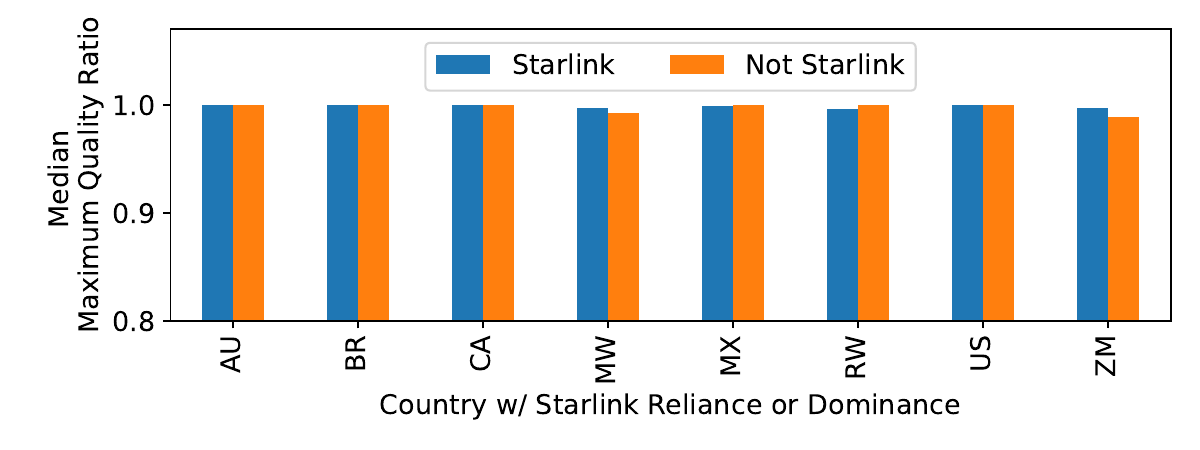}
        \caption{\textbf{Median Maximum Quality Ratio Per Country}---%
        \textnormal{In April~2024, the median Maximum Quality Ratio is high across countries with Malawi and Zambia experiencing a small improvement for sessions over Starlink when compared to non-Starlink networks.}}
        \label{fig:50q_MQR_country}
    \end{subfigure}%
    \hfill
    \begin{subfigure}[t]{0.49\textwidth}
        \centering
        \includegraphics[scale=0.35]{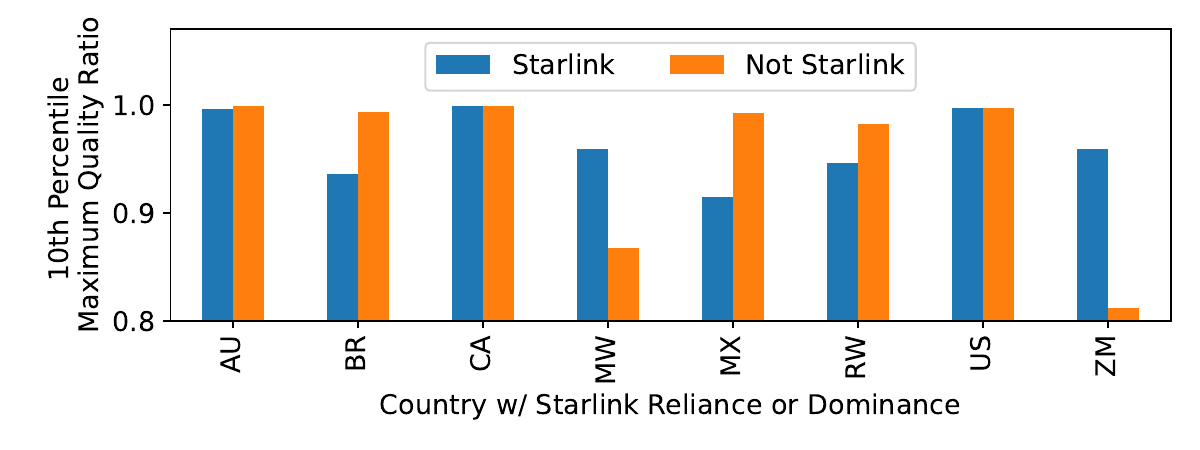}
        \caption{\textbf{Long-tail Maximum Quality Ratio Per Country}---%
        \textnormal{In April~2024, Starlink provides up to 18\% better Maximum Quality Ratio in the worst video streaming sessions in Malawi and Zambia, and up to 8\% worse in the worst streaming sessions across Mexico, Rwanda, and Brazil.}}
        \label{fig:10q_MQR_country}
    \end{subfigure}
    \caption{\textbf{Comparison of Perceptual Video Quality (Maximum Quality Ratio) Per Country for Starlink and Non-Starlink Networks}}
    \vspace{-15pt}
    \label{fig:combined_vmaf}
\end{figure*}

The vast majority (83\%) of countries served by Starlink experience a Maximum Quality Ratio that is equal to or, in seven countries, better than the median ratio provided by alternative ISPs. In Figure~\ref{fig:50q_MQR_country}, we plot the median Maximum Quality Ratio for Starlink and non-Starlink ISPs across all streaming sessions per country. This analysis focuses on countries that either dominate video streaming over Starlink or rely the most on Starlink for video streaming, as discussed in Section~\ref{sec:rely_dom_country}. Note that we exclude the Pitcairn and Christmas Islands due to their small sample sizes of video streaming sessions, and we do not compare against the top 10 ISPs, as they are generally unavailable in the majority of the selected countries.

The median Maximum Quality Ratio is above 0.98 across countries for both Starlink and non-Starlink networks. In Malawi and Zambia, Starlink sessions show a small improvement (less than 1\%) in the median Maximum Quality Ratio. More importantly, the perceptual quality improvement in Malawi and Zambia is significant for sessions with lower Maximum Quality Ratios (i.e., the lowest 10\% quality streams). Figure~\ref{fig:10q_MQR_country} illustrates the 10th percentile Maximum Quality Ratio for the same set of countries. Starlink provides a 10.6\% better Maximum Quality Ratio in the worst video streaming sessions in Malawi, and 18.1\% in Zambia. Consequently, it is no surprise that customers in Malawi and Zambia rely on Starlink for video streaming more than in other countries (Section~\ref{sec:rely_dom_country}). Conversely, Starlink provides up to 8\% worse Maximum Quality Ratios in the worst streaming sessions across Mexico, Rwanda, and Brazil.

\netflix's perspective adds a new dimension for how to reason about Starlink's network performance in Africa relative to prior work.
Prior work came to the conclusion that Starlink performance in Africa is strictly worse, due to dramatically higher latency~\cite{izhikevich2024democratizing}. 
However, our results show that even in light of such latency, perceptual video quality over Starlink can be better than local alternatives in select countries in Africa.


\begin{wrapfigure}{t}{0.5\textwidth}
  \vspace{-5pt}
    \includegraphics[scale=0.35]{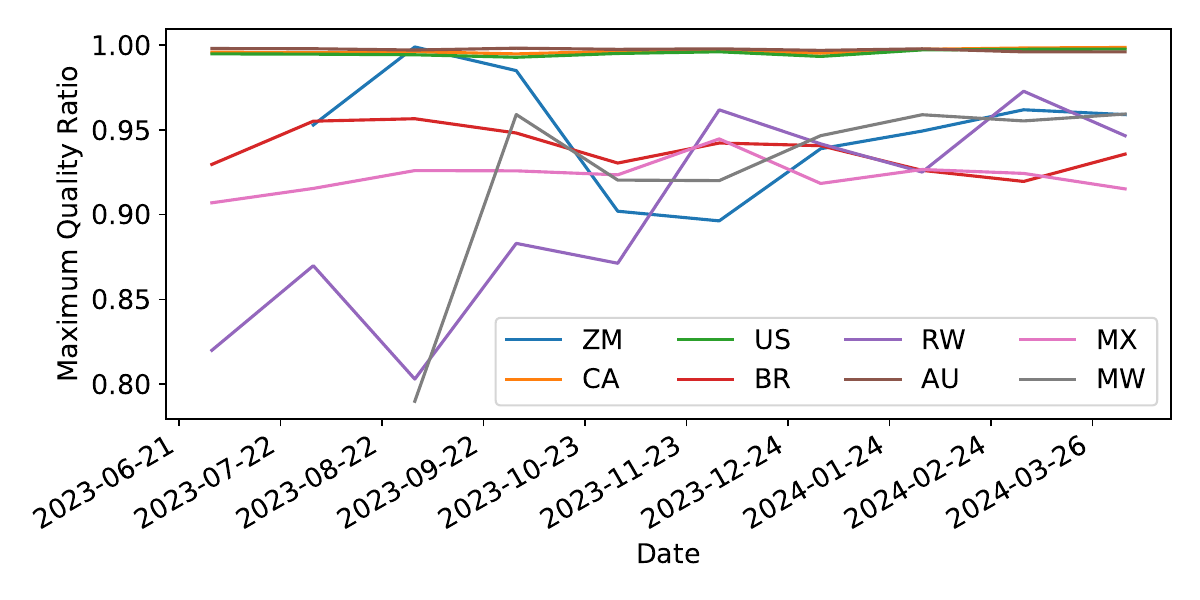}
    \caption{\textbf{Starlink Maximum Quality Ratio Country Trends}---%
    \textnormal{Over time, perceptual video quality over Starlink has improved. Regions that rely more on Starlink experience a more dramatic increase in video streaming quality over time. }}
    \vspace{-5pt}
    \label{fig:country_MQR_overtime}
\end{wrapfigure} 

Over the past year, the perceptual video quality of Starlink sessions has improved, approaching the maximum achievable quality for each session, especially among countries with room for improvement. In Figure~\ref{fig:country_MQR_overtime}, we plot the median Maximum Quality Ratio over time for countries that either dominate or heavily rely on Starlink for video streaming. The US, Canada, and Australia consistently achieve the best possible quality over Starlink, with a median Maximum Quality Ratio above 0.98 throughout the year. African countries, including Rwanda, and Malawi have shown the most significant improvement in median Maximum Quality Ratio over time, gradually increasing to approximately 0.95.
Interestingly, these African countries now achieve higher Maximum Quality Ratios than Latin American countries (Brazil and Mexico), which have fluctuated between 0.9 and 0.95 throughout the year.

\begin{figure*}[t]
    \vspace{-15pt} 
    \centering
    \begin{subfigure}[t]{0.48\textwidth} 
        \vtop{\vskip0pt 
        \hbox{\includegraphics[width=\linewidth]{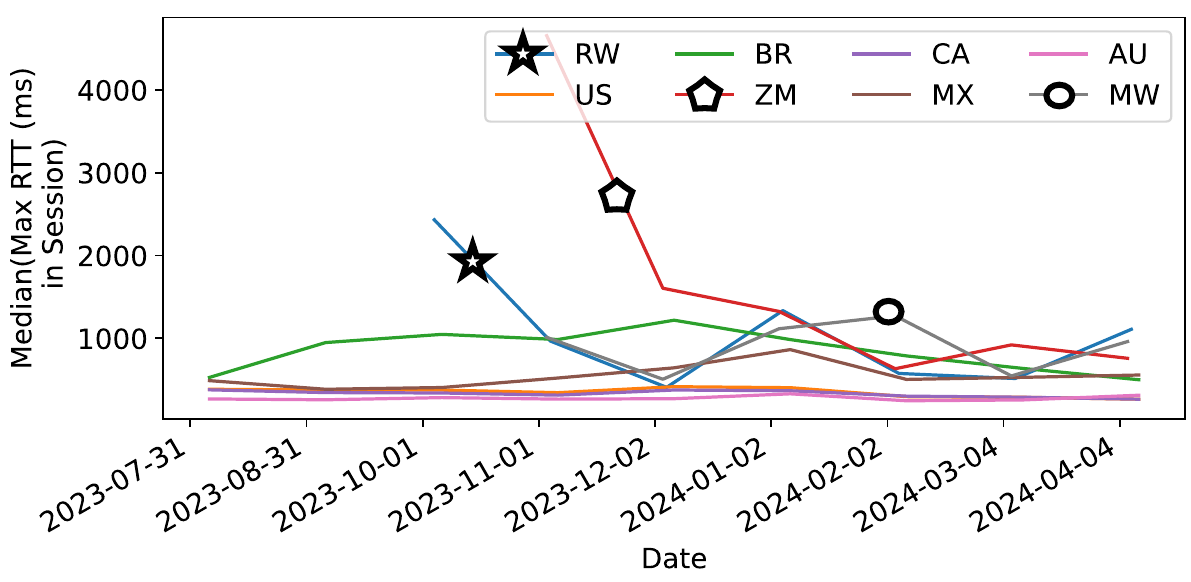}}}
        \caption{\textbf{Maximum RTT}}
        \label{fig:max_rtt_overtime}
    \end{subfigure}
    \hfill
    \begin{subfigure}[t]{0.48\textwidth} 
        \vtop{\vskip0pt 
        \hbox{\includegraphics[width=\linewidth]{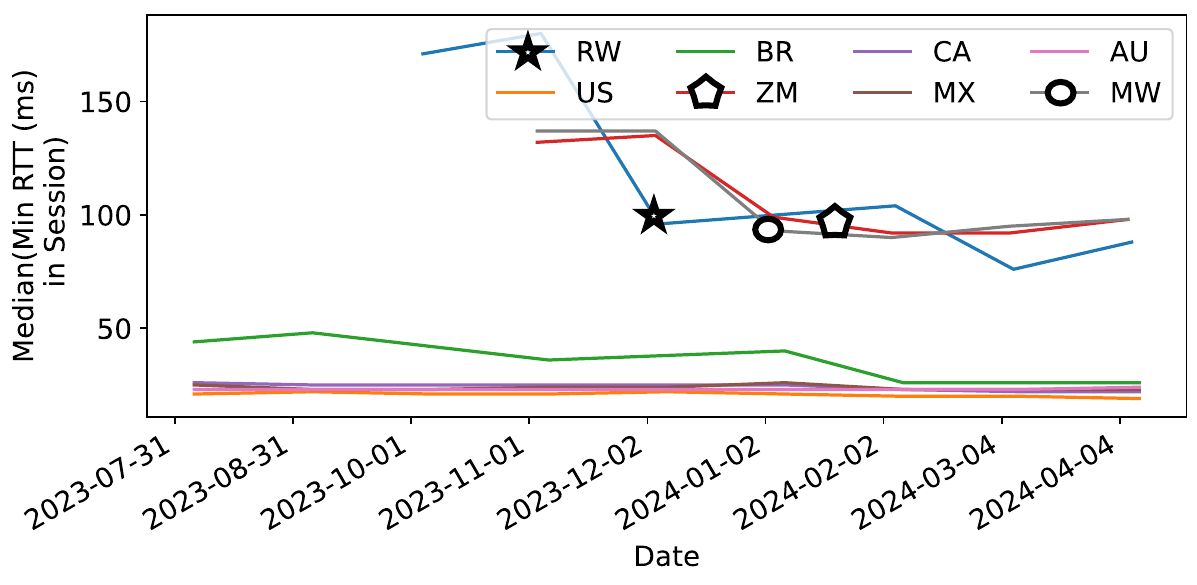}}}
        \caption{\textbf{Minimum RTT}}
        \label{fig:min_rtt_overtime}
    \end{subfigure}
    \caption{\textbf{RTT Over Time}---%
    \textnormal{Countries in Africa (Zambia (ZM), Rwanda (RW), Malawi (MW)) experience dramatic decreases in latency over time relative to other countries.}}
    \label{fig:rtt_overtime}
    \vspace{-15pt} 
\end{figure*}

Africa's VMAF improvement coincides with dramatic decreases in round trip latencies.
In Figure~\ref{fig:max_rtt_overtime} and Figure~\ref{fig:min_rtt_overtime} we plot the maximum and minimum round trip times (RTT), respectively, experienced during a streaming session over Starlink across different countries over time. 
Round trip times are calculated by continuously measuring the time between the sending of a TCP segment (e.g., with a video chunk) and the receipt of acknowledgment from the client. 
Countries in Africa experience the most dramatic minimum and maximum round trip times decreases over time.
For example, Rwanda's minimum round trip times decrease two-fold between October and December~2023.
Zambia's maximum round trip times decrease six-fold between October~2023 and February~2024. 
Starlink shares that decreased latency is due to increased ground
infrastructure, improved routing, and more satellites~\cite{nathan_latency_report}.
Nevertheless, minimum RTT across Africa remains nearly twice as high (Figure~\ref{fig:min_rtt_overtime}).
The excessively high RTT (i.e., occasionally on the order of seconds) could impact VMAF by creating too much delay when reacting to changes in network conditions. 

\label{sub:sec:rtt_overtime}

\begin{wrapfigure}{t}{0.5\textwidth}
  \vspace{-5pt}
  \includegraphics[width=\linewidth]{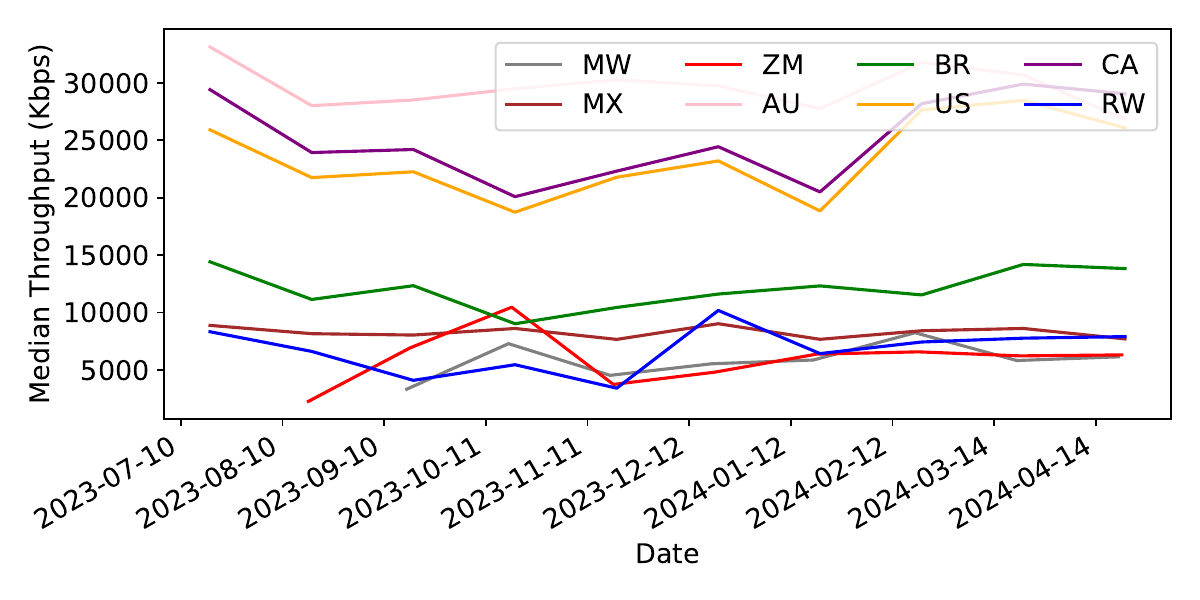}
  \caption{\textbf{Starlink Throughput Country Trends}---%
  \textnormal{Over time, Starlink throughput remains relatively stable, with Australia experiencing the highest throughput.}}
   \vspace{-20pt}
  \label{fig:country_throughput_overtime}
\end{wrapfigure}

Meanwhile, throughput over Starlink has remained relatively stable over time.
In Figure~\ref{fig:country_throughput_overtime}, we plot the throughput over Starlink over time.
Throughput is measured as the average of the number of bytes downloaded divided by the time spent for download.  
Australia consistently experiences the highest throughput, which may explain why Australian users experience the highest video perceptual quality (higher throughput often leads to higher bitrate selection).
Nevertheless, throughput does not appear to contribute to Africa's video perceptual quality improvement, as Africa's throughput remains relatively stable over time. 

\subsection{Play Delay}

\begin{figure*}[t] 
    \vspace{-15pt} 
    \centering
    \begin{subfigure}[t]{0.48\textwidth} 
        \includegraphics[width=\linewidth]{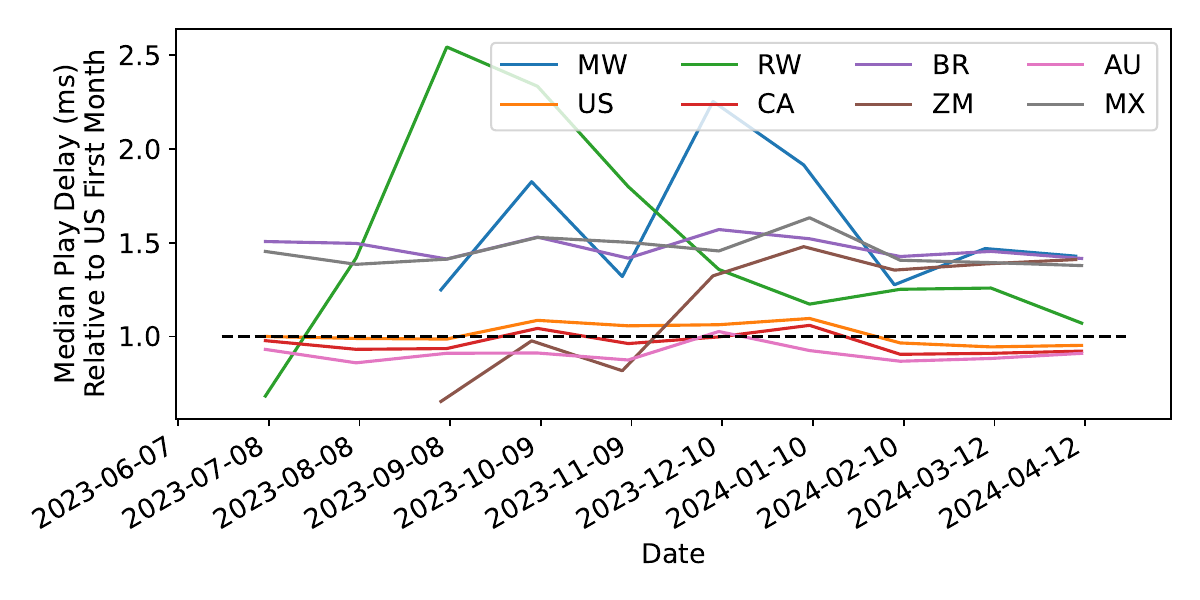}
        \caption{\textbf{Starlink Play Delay Country Trends}---%
        \textnormal{Over time, play delay over Starlink has improved. Regions that rely more on Starlink experience a more dramatic decrease in play delay over time.}}
        \label{fig:country_playdelay_overtime}
    \end{subfigure}
    \hfill
    \begin{subfigure}[t]{0.48\textwidth} 
        \includegraphics[width=\linewidth]{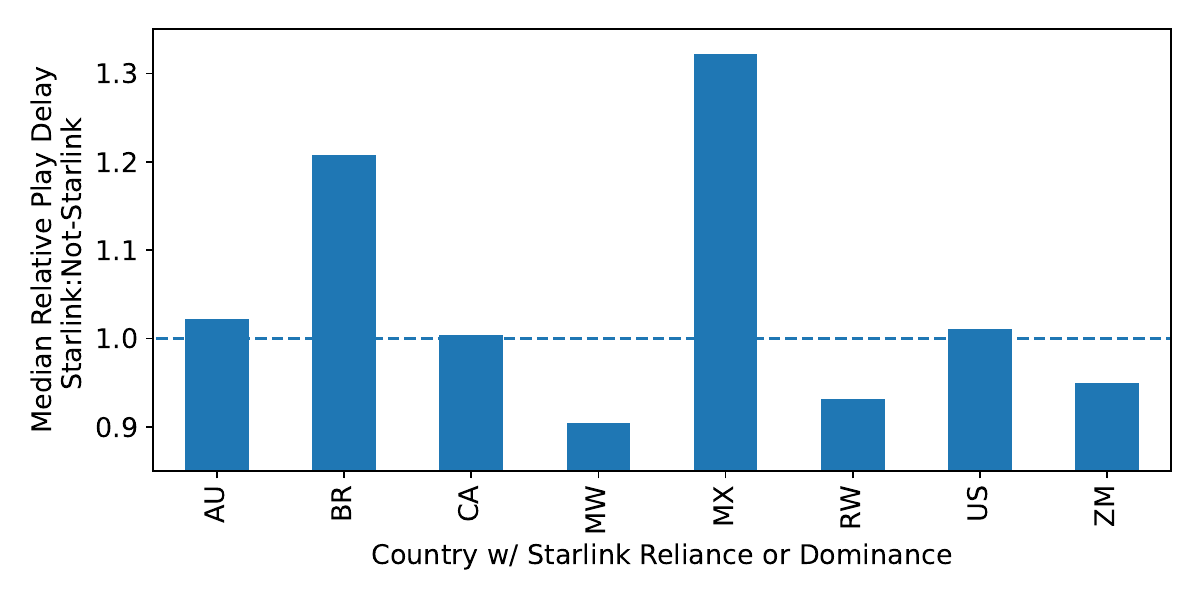}
        \caption{\textbf{Play Delay Per Country}---%
        \textnormal{In April~2024, countries in Africa (Zambia (ZM), Rwanda (RW), Malawi (MW)) experience significantly decreased play delays over Starlink relative to non-Starlink alternatives.}}
        \label{fig:country_playdelay_relative}
    \end{subfigure}
    \caption{\textbf{Play Delay Trends and Comparisons}}
    \label{fig:play_delay_comparison}
    \vspace{-15pt} 
\end{figure*}

Africa's higher RTTs also lead to higher play delays (i.e., the time between when video is first requested and when it actually begins playing), which contributes negatively to overall quality of experience over Starlink.
We calculate the median play delay across every country for every month in the past year, and plot the evolution of play delay in Figure~\ref{fig:country_playdelay_overtime} (normalized by the play delay of the US in the first month of the year). 
Countries with historically high RTTs (Zambia, Rwanda, Malawi) experienced  play delays that were up to 250\% higher than the US. 
However, due to their RTTs decreasing over time (Figure~\ref{fig:rtt_overtime}), play delays also decreased. 

While play delays across Africa, as of April~2024, remain at least 15\% higher than other parts of the world, their play delay is better than non-Starlink alternatives.
We calculate the median play delay per country in April~2024, and plot the ratio between Starlink and not Starlink's play delay in Figure~\ref{fig:country_playdelay_relative}.
Zambia, Rwanda, Malawi experience play delays that are 5--10\% lower than non-Starlink alternatives. 
\looseness=-1

\subsection{Network Impact on Video Delivery}

While Starlink users experience overall perceptual video quality similar to non-Starlink networks, they are more likely to experience bitrate switches and network rebuffers due to Starlink's network conditions. 
Our results in this section are what lead us to eventually manipulate existing congestion control (Section~\ref{sec:tcp}) and adaptive bitrate streaming algorithms (Section~\ref{sec:abr_design}).

\label{sec:sub:sub:bitrate_switches}
\begin{wrapfigure}{t}{0.5\textwidth}
  \includegraphics[width=\linewidth]{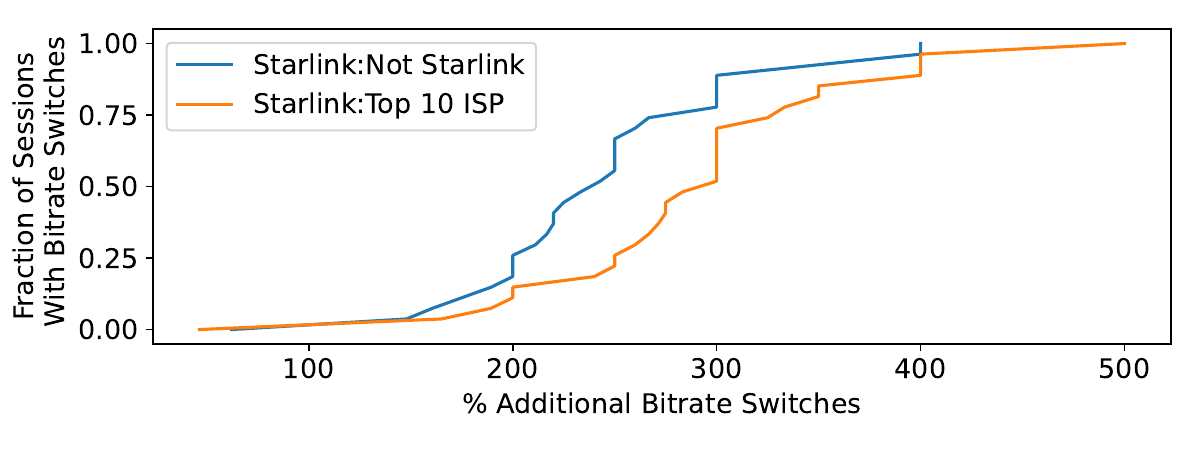}
  \caption{\textbf{Bitrate Switches}---%
  \textnormal{Video streaming over Starlink suffers from increased bitrate switches relative to other networks in April~2024.}}
  \vspace{-10pt}
  \label{fig:bitrate_switch}
\end{wrapfigure}

\subsubsection{Bitrate Switches}


Starlink users are significantly\footnote{We use a one-sided Kolmogorov-Smirnov test with $p < 0.05$.\label{foot:ks}} (60\%) more likely to experience a bitrate switch while streaming video over \netflix, which contributes negatively to quality of experience.
About half of sessions with a bitrate switch experience a resolution of less than 1080p.
In Figure~\ref{fig:bitrate_switch}, we plot the distribution of how many additional bitrate switches a Starlink session experiences relative to other networks. 
For sessions that experience at least one bitrate switch, 50\% of Starlink sessions will experience over 2~times the number of bitrate switches compared to non-Starlink networks. 

\begin{wrapfigure}{t}{0.5\textwidth}
\vspace{-10pt}
  \includegraphics[width=\linewidth]{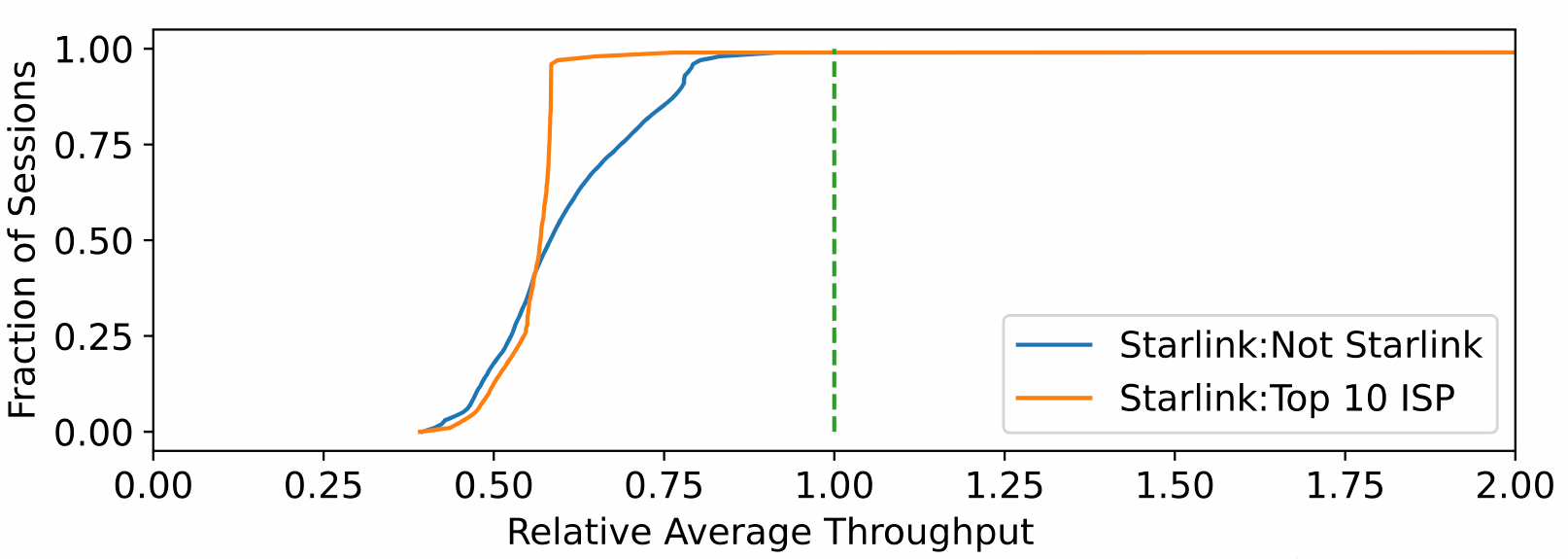}
  \caption{\textbf{Throughput Distribution}---%
  \textnormal{Starlink users experience lower throughput while streaming over a single TCP connection in April~2024. The ratio of throughput is calculated by dividing quantiles of measurements.}}
  \vspace{-10pt}
  \label{fig:throughput_dist}
\end{wrapfigure}

\label{sub:sec:throughput_less}
Starlink's bitrate switches are likely due to Starlink's lower and oscillating throughput.
In Figure~\ref{fig:throughput_dist}, we plot the distribution of relative average throughput recorded from the clients receiving video.
Uniquely, over 95\% of Starlink's throughput is lower than alternative networks. 
For example, Starlink throughput is nearly always 50\% of what a top 10 ISP offers and falls below 20~Mb/s---a bandwidth that is often necessary if streaming  4K video~\cite{youtube_bitrates}.
Over 90\% of bitrate switches, no matter the network, occur at throughputs below 20~Mb/s.

\begin{figure*}[t]
    \vspace{-10pt} 
    \centering
    \begin{subfigure}[t]{0.48\textwidth} 
        \vtop{\vskip0pt 
        \hbox{\includegraphics[width=\linewidth]{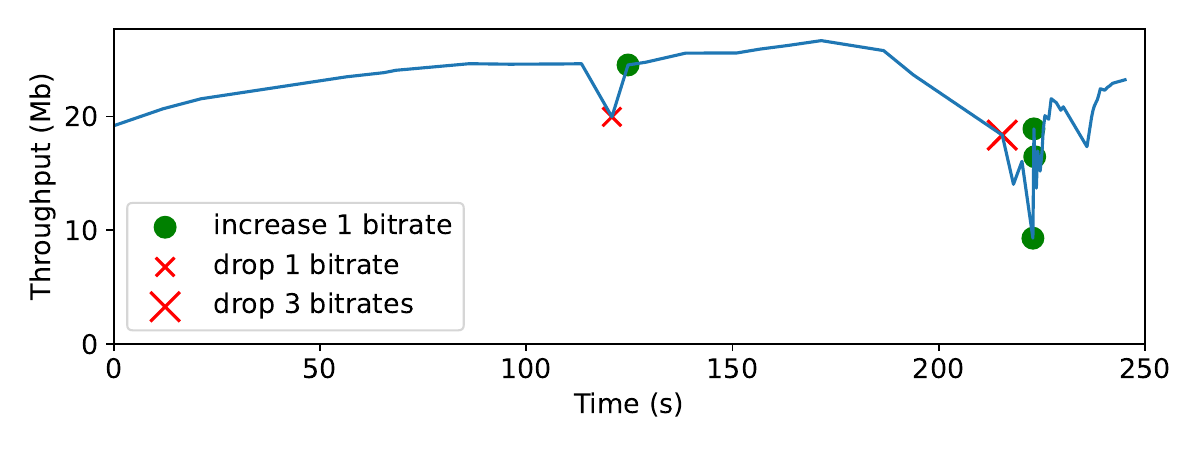}}}
        \caption{\textbf{Example Session Throughput With Associated Bitrate Switches}}
        \label{fig:throughput_bitrates}
    \end{subfigure}
    \hfill
    \begin{subfigure}[t]{0.48\textwidth} 
        \vtop{\vskip0pt 
        \hbox{\includegraphics[width=\linewidth]{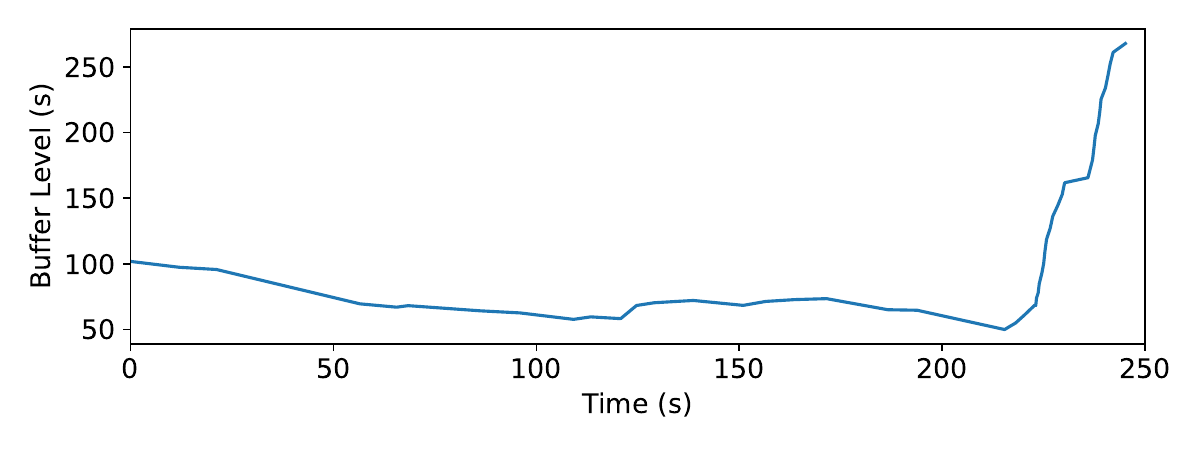}}}
        \caption{\textbf{Example Session Buffer Level}}
        \label{fig:throughput_buffer}
    \end{subfigure}
    \caption{\textbf{Starlink Bitrate Switches}---%
    \textnormal{Video streaming sessions over Starlink that suffer from bitrate switches are often due to Starlink's lower and oscillating throughput.}}
    \label{fig:ex_bitrate_switch}
    \vspace{-10pt} 
\end{figure*}

\label{sub:sec:variable_throughput}
Furthermore, Starlink throughput is more heavily driven by variance; 80\% of observed throughputs increase/decrease with greater magnitude than in a Top 10 ISP. 
On average, Starlink throughput increases/decreases take longer to recover:
While 50\% of throughput decreases that reach below 10Mb/s---a bandwidth that is often necessary if streaming  1080p video~\cite{youtube_bitrates}---in non Starlink networks take roughly 5 seconds to recover back to above 10Mb/s, 50\% of Starlink throughput decreases take roughly 15 seconds to recover.
Coincidentally, 15 seconds is the interval that Starlink re-configures static routes for packet forwarding between satellites and/or ground stations~\cite{starlink_fcc}.

In Figure~\ref{fig:throughput_bitrates}, we illustrate an example video streaming session over Starlink, which suffers from a total of eight bitrate switches within the span of 100~seconds. 
The Starlink throughput dipping below 20~Mb/s twice while the buffer level dips below holding less than 1~minute of video (Figure~\ref{fig:throughput_buffer}) contributes to the decreases in bitrates. 
Meanwhile, a sudden upward burst of throughput contributes to the successive increases in bitrate. 

\begin{wraptable}{t}{0.5\textwidth}
\centering
\small
\begin{tabular}{llll}
\toprule

Region & Likelihood of >= 1 Rebuffer\\
 & Relative to US\\
	\midrule  
Canada & 0.66x \\
Asia Pacific & 0.9x \\
United States & 1x \\
Europe & 1.07x \\
Africa & 2.7x \\
Latin America & 3.7x \\
\bottomrule \end{tabular}
\vspace{5pt}
	\caption{\textbf{Regional Rebuffer Likelihood Over Starlink}---\textnormal{Starlink streamers near Canada experience the smallest likelihood of experiencing at least one rebuffer. Meanwhile under-served regions in Africa, and Latin America, are significantly more likely to experience at least one rebuffer. }}
 \vspace{-20pt}
\label{table:rebuffer_regions}
\end{wraptable}

Bitrate switches, while playing a key role in ensuring that no rebuffers occur, still negatively affect video streaming quality of experience.
While overall time weighted perceptual video quality (VMAF) is not affected by the bitrate switches, as shown in Section~\ref{sub:sec:vmaf}, time weighted VMAF only captures an aggregated view of session quality over the full duration of a session (which can be more than one hour for a movie). It does not capture differences in quality between video segments within a session.
For example, Ho{\ss}feld et al.~\cite{hossfeld2014assessing} find that the amplitude and frequency of bitrate switches affect overall quality of experience (e.g., constant and large quality switching can cause more irritation than sticking to one quality level). 
In fact, a community of Starlink users complain on Reddit about excessive bitrate switching during video streaming~\cite{reddit_bitrate_switches}.
In Section~\ref{sec:tcp}, we show how the congestion controller can be manipulated to improve throughput and decrease bitrate switches, while not significantly increasing network rebuffers. 
\looseness=-1

\subsubsection{Network Rebuffers}
\label{sub:sub:sec:net_rebuffers}


Video rebuffers are one of the most severe contributors to a negative quality of experience~\cite{huang2014buffer,dobrian2011understanding}.
While video rebuffers due to network conditions are extremely rare across all networks, they are significantly~\footref{foot:ks} (216\%) more likely to occur over Starlink than a Top~10 ISP, and 40\% more likely to occur relative to any non-Starlink network. 
Furthermore, if a customer experiences at least one rebuffer, Starlink customers are 50\% more likely to experience more than one rebuffer.
In the worst case, Starlink customers experience twice as many rebuffers as non-Starlink customers.

\begin{figure*}[t]
    \vspace{-10pt} 
    \centering
    \begin{subfigure}[t]{0.48\textwidth} 
        \vtop{\vskip0pt 
        \hbox{\includegraphics[width=\linewidth]{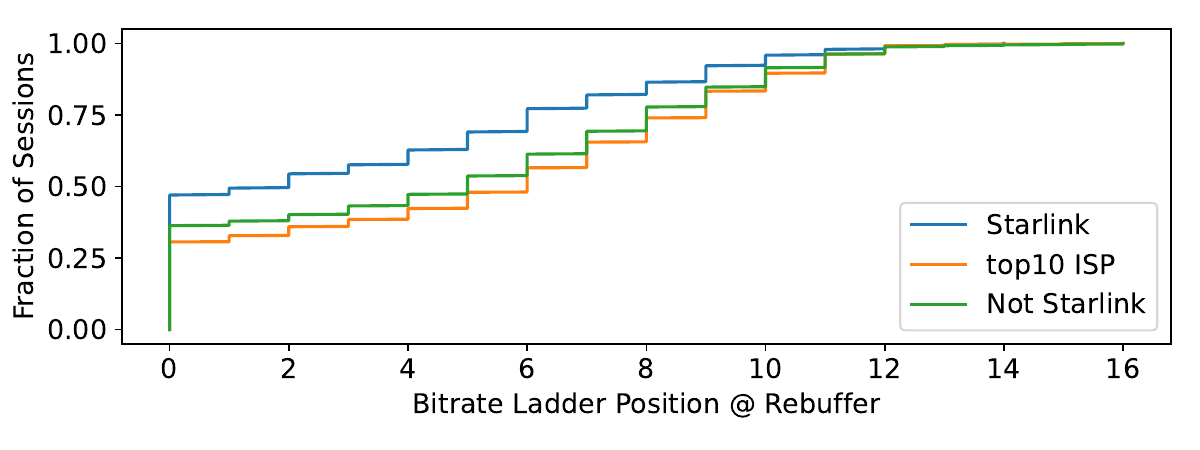}}}
        \caption{\textbf{Bitrate Ladder Position At The Rebuffer}---%
        \textnormal{Streaming sessions over Starlink are more likely to already be at a video's worst quality before a rebuffer occurs. The bitrate ladder denotes what bitrate rank (0 being the lowest) is chosen at the time of a rebuffer.}}
        \label{fig:bitrate_at_rebuffer}
    \end{subfigure}
    \hfill
    \begin{subfigure}[t]{0.48\textwidth} 
        \vtop{\vskip0pt 
        \hbox{\includegraphics[width=\linewidth]{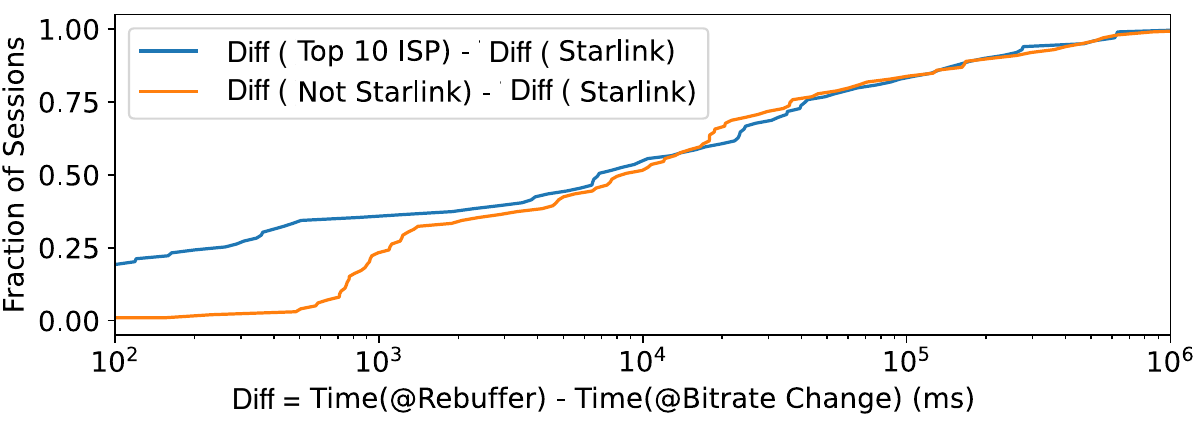}}}
        \caption{\textbf{Time to Downshift Bitrate Before Rebuffer}---%
        \textnormal{Streaming sessions over Starlink spend less time in the reduced bitrate state before a rebuffer occurs.}}
        \label{fig:time_to_react_to_rebuffer}
    \end{subfigure}
    \caption{\textbf{Starlink Streaming Performance: Bitrate and Rebuffering}}
    \label{fig:combined_bitrate_rebuffer}
    \vspace{-10pt} 
\end{figure*}

Rebuffers over Starlink are more likely to occur in historically under-connected regions. 
For every region, we calculate the percent of streaming sessions that experience at least on rebuffer, denoted as \textit{rebuffer\_sessions}. 
In Table~\ref{table:rebuffer_regions}, for every region we divide its \textit{rebuffer\_sessions} by the US's \textit{rebuffer\_sessions}.
Streaming sessions near Africa and Latin America are over 2~times more likely to experience a rebuffer relative to the United States.

Rebuffers over Starlink are likely exacerbated due to (1)~lower throughputs, causing even the lowest bitrate selection to become a bottleneck  and (2)~throughputs with high variance, causing adaptive bitrate algorithms to not have enough time or signal to downshift bitrates before it is too late. 
In Figure~\ref{fig:bitrate_at_rebuffer}, we plot the distribution of what bitrate rank (0 being the lowest) is chosen at the time of a rebuffer. 
Sessions over Starlink are 60\% more likely to already be serving the lowest supported bitrate at the time of a rebuffer, indicating that an optimal bitrate selection is not to blame for the rebuffers.

Rather, streaming sessions over Starlink spend less time in reduced bitrate state, not allowing for enough time for the buffer to be filled, thereby increasing the chances of a rebuffer. 
In Figure~\ref{fig:time_to_react_to_rebuffer}, we plot the distribution of the difference in time between when a rebuffer occurs and when the last bitrate change occurs. 
In the median case, Starlink sessions spend 50\% less time in the reduced bitrate state. 
Limiting the amount of time Starlink sessions fill their buffer at the reduced bitrate hurts Starlink sessions especially because Starlink sessions are six times more likely to experience an outage (i.e., a period of at least 8~seconds\footnote{The 8~second threshold is not significant, but rather arises from the manner in which data is collected.} compared to non Starlink sessions. 
Outages are often due to a physically obstructed dish or no satellite in line of sight~\cite{tanveer2023making}.

In Section~\ref{sec:abr_design}, we show how designing adaptive bitrate selection algorithms that are more reactive to throughput changes can reduce the chances of a rebuffer occurring.

\section{Improving Congestion Control for Video Delivery over LEO}
\label{sec:tcp}

In Section~\ref{sec:sub:sub:bitrate_switches}, we found that Starlink users experience lower throughput than users on other ISPs, which leads to increased bitrate switches and network rebuffers. 
In this section, we explore the potential causes and mitigations for low throughput by focusing on the algorithm that determines the average rate at which the server sends data: congestion control. 
Through deploying A/B tests on over 1~million \netflix users, we discover that while we are able to modify the congestion controller to achieve higher throughput and fewer bitrate switches, it comes at an undesirable cost of increased re-transmit rates and latency.
We finish with a discussion on choices of congestion controllers to account for a diverse set of network characteristics. 

\vspace{3pt}
\noindent
\textbf{Ethics.} \quad
Our A/B experiments involve minor tweaks to the configuration of \netflix's transport mechanisms. These experiments are designed to impact only a small fraction of users and are part of \netflix's standard practice of rigorously testing every service change to continually improve user experience. \netflix conducts these tests under strict ethical guidelines, ensuring that all participants have the ability to opt out at any time. 
Furthermore, \netflix ensures that all experiments are conducted responsibly. Any potential adverse effects on user experience are closely monitored, and measures are in place to mitigate any negative impacts, including the ability to quickly stop any test that has an unintended adverse effect.

\subsection{Investigating Congestion Control Causes for Low Throughput}


\label{sec:sub:sub:rtt}
\begin{wrapfigure}{r}{0.5\textwidth} 
    \vspace{-10pt} 
    \includegraphics[width=\linewidth]{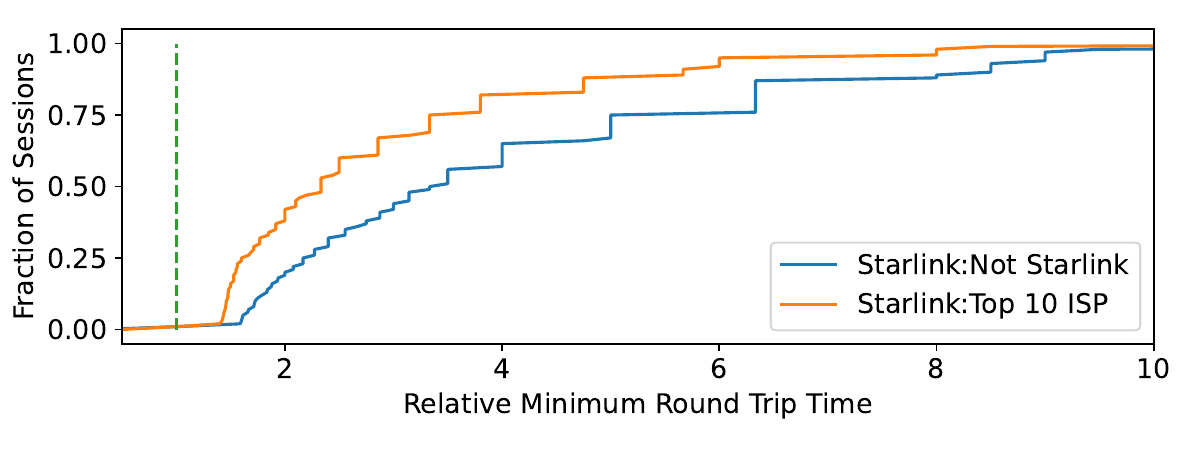}
    \caption{\textbf{Starlink's Minimum RTT Compared to Other Networks}---%
    \textnormal{Starlink sessions have longer minimum Round Trip Times than others. Relative RTT is calculated by dividing quantiles of RTT. For example, if the first percentile of all Starlink sessions experience an RTT of 100ms and the first percentile of all non-Starlink sessions experience an RTT of 50ms, then the resulting ratio would be $100/50$ (i.e., the first percentile of Starlink sessions experience 2x the RTT of non-Starlink's RTT). }}
    \label{fig:rtt_dist}
\end{wrapfigure}

Starlink's high packet round trip times (RTTs) and increased retransmit rates contribute to low TCP congestion windows, and hence, lower throughput.  
We find (Figure~\ref{fig:rtt_dist}) that 75\% of minimum RTTs over Starlink are at least two times greater than non-Starlink networks. 
The higher latency is consistent with prior work, which attributes the latency to longer routing paths~\cite{izhikevich2024democratizing}.
Further, we discover the majority of Starlink sessions experience nearly two times the retransmit rate relative to non-Starlink networks.
The increase in Starlink retransmits is likely influenced by random loss on satellite links~\cite{mohan2023multifaceted}; the use of Active Queue Management (FQ-CoDel) on the Starlink WiFi router~\cite{nathan_latency_report}, which minimizes latency at the expense of packet loss; and potential packet reordering~\cite{hypatia}, which may lead to duplicate acknowledgments and spurious retransmits.

\begin{wrapfigure}{t}{0.5\textwidth} 
    \includegraphics[width=\linewidth]{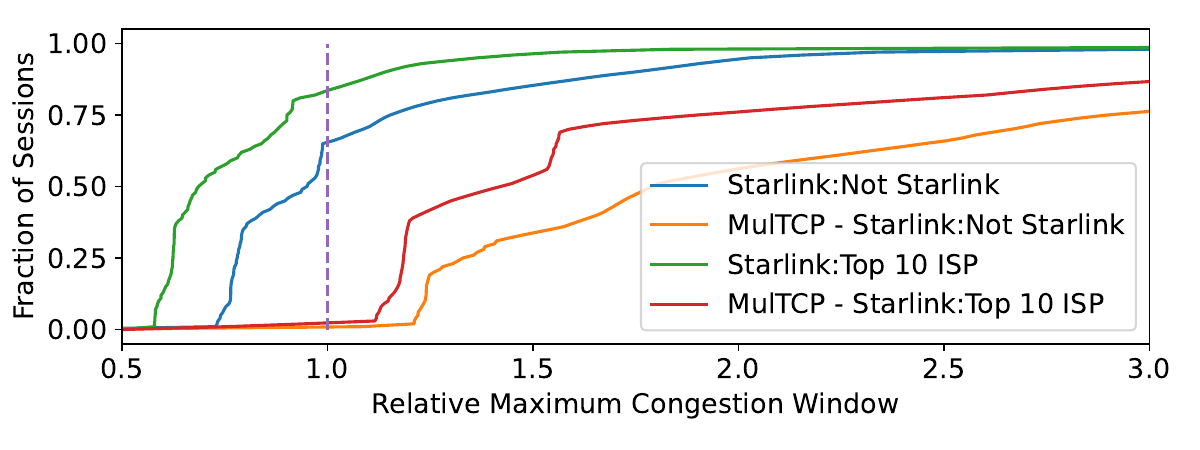}
    \caption{\textbf{Congestion Windows Across Networks}---%
    \textnormal{The majority of Starlink sessions have lower congestion windows than non-Starlink networks (i.e., the majority of values in both distributions are less than 1), but emulating an increase in concurrent TCP connections increases congestion windows well beyond non-Starlink alternatives (i.e., the majority of values in both distributions are greater than 1). }}
    \label{fig:cwnd_ab}
\end{wrapfigure}

Increases in RTTs and retransmit rates can cause loss-based congestion controllers---based on Additive Increase Multiplicative Decrease (AIMD)---to take a long time to grow the TCP congestion window~\cite{garcia2023multi}. 
In addition, loss-based congestion controllers assume that any packet loss is due to congestion in the network, and therefore, back off when seeing indications of packet loss. 
To confirm that the reduced throughput is correlated with a reduced ability of TCP to grow its congestion window, we compare the maximum congestion windows achieved by connections streaming over Starlink to those streaming over other networks, focusing on large-enough connections that transfer at least 10MB. 
Indeed, when using the default-on congestion controller that \netflix currently uses, New Reno~\cite{floyd1999tcp}, we see evidence of congestion windows for \netflix sessions over Starlink paths being lower than over other networks (see lines labeled ``Starlink'' in Figure~\ref{fig:cwnd_ab}).
With lower congestion windows, Starlink sessions are more likely to underutilize the true available bandwidth for a single TCP connection, and therefore experience lower throughput. 

\subsection{Experiment Setup: New Reno with MulTCP}

Single TCP connections are unable to fully utilize the Starlink path; thus, we develop a modified congestion control with the goal of improving throughput and overall video quality of experience.
Multiple concurrent connections may be able to utilize a larger fraction of the available capacity on the path, but we cannot easily change the number of parallel connections that the client uses.
To approximate the use of multiple connections, we create a modified version of New Reno with MulTCP~\cite{crowcroft1998differentiated} that emulates the behavior of three concurrent connections: it grows the congestion window by three bytes for every acknowledged byte, and when it detects packet loss, it reduces the congestion window as if only one of the three emulated flows had seen a loss event.
Thus, New Reno with MulTCP is able to grow the congestion window faster and does not back off as much as New Reno on paths with long RTTs and unsynchronized or random packet loss. 
We do not experiment with LEO-specific TCP replacements, as they would likely not be deployed at an environment like \netflix, where hundreds of millions of users rely on a myriad of types of access networks and dynamically choosing network-specific TCP replacements would introduce significant operational complexity.

We run an A/B test using New Reno with MulTCP on over 1~million \netflix video streaming sessions over Starlink.
The A/B test randomly assigns a small fraction of Starlink's users to either the control group (which received the default congestion control treatment) or the treatment group (which received the modified congestion controller treatment).
We run the test for a week in mid-April~2024.

\subsection{Results}

\begin{wrapfigure}{t}{0.5\textwidth} 
    \vspace{-10pt} 
    \includegraphics[width=\linewidth]{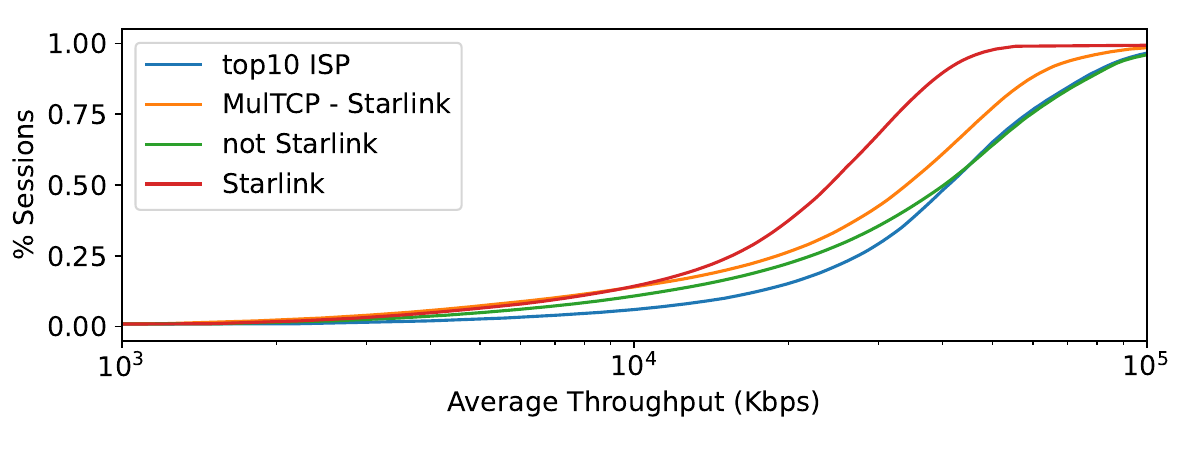}
    \caption{\textbf{MulTCP Effect on Starlink Throughput}---%
    \textnormal{Emulating an increase in concurrent TCP connections increases average throughput over Starlink to be near non-Starlink alternatives.}}
    \label{fig:throughput_ab}
    \vspace{-10pt} 
\end{wrapfigure}

Emulating an increase in concurrent TCP connections (i.e., New Reno with MulTCP) increases average throughput over Starlink to be near non-Starlink alternatives.
In Figure~\ref{fig:throughput_ab}, we plot the distribution of throughput of our control group (Starlink), our treatment group (MulTCP - Starlink), and non-Starlink alternatives. 
New Reno with MulTCP increases average throughput by 30-40\% compared to the default congestion controller, and it increases the congestion window (Figure~\ref{fig:cwnd_ab}).

New Reno with MulTCP improves overall video quality of experience. 
Increase in throughput results in a statistically significant,\footnote{We use a Mann-Whitney test comparing the quantile distribution of the chosen metric to derive these statistically significant results.} 0.08\%--1\% increase in lower quantile VMAF.
Further, network rebuffers per hour of video played decrease by 7\%, play delay decreases by 2\%, and bitrate switches decrease by 5\%.
Nevertheless, New Reno with MulTCP over Starlink is still 33\% more likely than non-Starlink sessions to experience a bitrate switch.

Unfortunately,  New Reno with MulTCP presents undesirable drawbacks.
Retransmit rates increase by 300\% on average and the 95th percentile of latency measurements increases by 77\% on average, indicating that  New Reno with MulTCP fills in-network queues to a greater capacity. 
Further, the increase in latency under load could cascade into other problems when competing with real-time flows.


Thus, while increasing throughput using  New Reno with MulTCP improves video streaming quality, it does so at the expense of increased re-transmission and latency.
As loss-based congestion controllers may not be able to fully utilize the Starlink path, future work should explore the use of a rate-based congestion controller such as BBR, that is based on path capacity measurements like delivery rate or goodput and that does not interpret every loss event as congestion.
However, given prior work evaluating the different versions of BBR~\cite{zeynali2024promises,yang2022bbrv2}, its use may come with its own tradeoffs as well, and further work may be needed to ensure a congestion controller works well across a wide range of paths including LEO networks.



\section{Adaptive Bitrate Streaming For Starlink}
\label{sec:abr_design}

In this section, we investigate how different design principles of adaptive bitrate streaming (ABR) help mitigate the increased rate of rebuffers over Starlink (Section~\ref{sub:sub:sec:net_rebuffers}) while achieving the highest quality possible.
We find that common ABR design is less effective at handling Starlink's variable throughput, often causing increased rebuffers or decreased quality relative to non-Starlink sessions. 
We discuss how future ABR designs should incorporate variance in throughput during bitrate selection. 

\subsection{Modeling ABR Today}
\label{sub:sec:abr_outline}
Adaptive bitrate streaming's fundamental goal is to select a video bitrate that achieves the highest possible quality while avoiding rebuffers.
Unfortunately, maximizing video quality is often at odds with minimizing rebuffers.
Given a steady network capacity, increasing video quality requires sending more data, causing video buffers to fill up slowly. 
Decreasing rebuffers---under variable throughput and availability---requires a larger video buffer, to account for the moments when sufficient throughput is not available.
ABR algorithms balance these opposing forces using a combination of two main inputs~\cite{yin2015control}: the throughput achieved and the buffer occupancy observed while downloading the previous chunks. 

In our experimentation of ABR over Starlink, we use an ABR algorithm that has three parameters that we can use to adjust the ABR's bitrate selection strategies in the face of throughput uncertainty -- \textit{throughput discount}, \textit{buffer discount}, and \textit{throughput smoothing}. At a high level,\footnote{Due to proprietary constraints, we are required to omit the actual ABR algorithm.} 
this algorithm follows a simple Model Predictive Control Style ABR model~\cite{sun2016cs2p} that picks the highest sustainable bitrate for chunk $i$ (denoted as $r_i$) such that $r_i \leq f(T_i, B_{i-1})$, where $T_i$ represents the estimated throughput for downloading chunk $i$ and $B_{i-1}$ represents the buffer occupancy after the download of the previous chunk $i-1$. 
In Section~\ref{sub:sec:results_sim}, we confirm that real deployed A/B tests reflect a subset of our simulated results. 

By tuning the throughput and buffer discount, we can make our ABR give more emphasis to the buffer occupancy as in buffer-based ABRs, to the throughput as in rate-based ABRs, or a mix of the two as in hybrid approaches. The \textit{throughput discount}, set as a percentage lower than 100\% of the  estimated throughput, accounts for uncertainties in throughput estimation. The \textit{buffer discount}, expressed in milliseconds, defines the buffer occupancy threshold below which the algorithm applies a greater discount to the throughput estimate. This adjustment is crucial when the buffer is nearly empty, as selecting lower bitrates can accelerate buffer refilling.  The \textit{throughput smoothing} parameter determines how to summarize past throughput observations, which throughput estimates are based upon. In our ABR experiments, we assume a simple estimator that uses an Exponentially Weighted Moving Average (EWMA) and use \textit{throughput smoothing} to control the half-life of the EWMA. Note that these three parameters are also commonly observed in many other ABRs in the literature~\cite{yin2015control,sun2016cs2p,yan2020learning}, and we treat them as building blocks of modern ABR algorithms.

\subsection{Methodology}
To understand how different ABR strategies mitigate rebuffers over Starlink and non-Starlink sessions, we simulate different ABR strategies on a random sample of 500K Starlink and 500K non-Starlink streaming sessions, randomly sampled between April~20--May 5, 2024. 
\netflix operates its own trace-driven simulator. For a sampled session, the simulator uses the session's previously-recorded throughput over time to simulate the network behavior. \netflix's player records throughput over time in 500~ms buckets, where throughput is computed as the number of bytes downloaded during a given bucket divided by the active transfer time during the bucket. The simulator then invokes \netflix's production ABR (outlined in Section~\ref{sub:sec:abr_outline}) over the recorded throughput to simulate the streaming of the sampled session.

\begin{wrapfigure}{t}{0.5\textwidth} 
    \includegraphics[width=\linewidth]{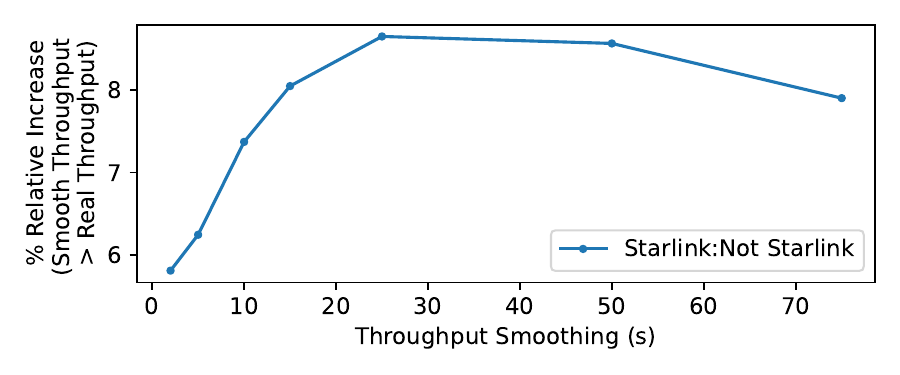}
    \caption{\textbf{Smoothing Throughput Overestimates Real Throughput}---%
    \textnormal{When Starlink throughput is smoothed, the smoothed throughput is up to 10\% more likely to overestimate the real throughput relative to non-Starlink throughputs. The y-axis denotes the relative error when overestimating the real throughput. We derive the y-axis by calculating (1) \% of video chunks across all sessions where the smoothed throughput is greater than the real throughput, denoted by $increase$, (2) the average across all video chunks: $avg(increase)$, and (3) the ratio of increase between Starlink and not Starlink sessions $Star(avg(increase))/NotStar(avg(increase))$.}}
     \vspace{-10pt}
    \label{fig:smoothin_overestimate}
\end{wrapfigure}

For each simulated ABR strategy, we vary one of three parameters while holding the other two at a constant value. More specifically, we sweep throughput smoothing window size (using exponentially weighted moving average) between 0--300000~ms, throughput discounts between 0--60\%, buffer level between 0--100000~ms at which to apply buffer discount.\footnote{If the buffer level is below \textit{buffer\_level}, we apply a throughput discount near 50\%, otherwise a throughput discount of less than 5\%.} 
When holding the parameter values constant, we use the following values:
throughput smoothing window near 100~seconds, buffer level near 50~seconds, and throughput discount near 15\%.
We derive the values by conducting an apriori simulation experiment---similar to the one described earlier---that chooses random values for all three parameters across 50~trials and searches for the combination of parameter values that balances the rebuffers and VMAF performance metrics.



\subsection{Results}

\label{sub:sec:results_sim}
Smoothing throughput, discounting throughput prediction, and discounting bitrates at different buffer levels are less effective at mitigating rebuffers for Starlink sessions during simulation and real deployed A/B tests. 

\begin{figure*}[t] 
    \vspace{-10pt} 
    \centering
    \begin{subfigure}[t]{0.48\textwidth} 
        \vtop{\vskip0pt 
        \hbox{\includegraphics[width=\linewidth]{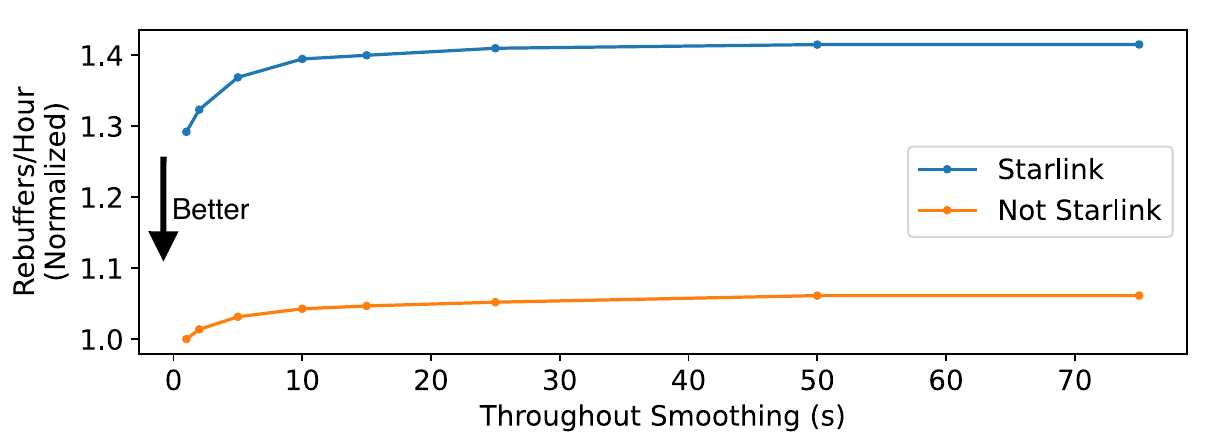}}}
        \caption{\textbf{Smoothing Effect On Network Rebuffers}}
        \label{fig:smooth_rebuffer}
    \end{subfigure}
    \hfill
    \begin{subfigure}[t]{0.48\textwidth} 
        \vtop{\vskip0pt 
        \hbox{\includegraphics[width=\linewidth]{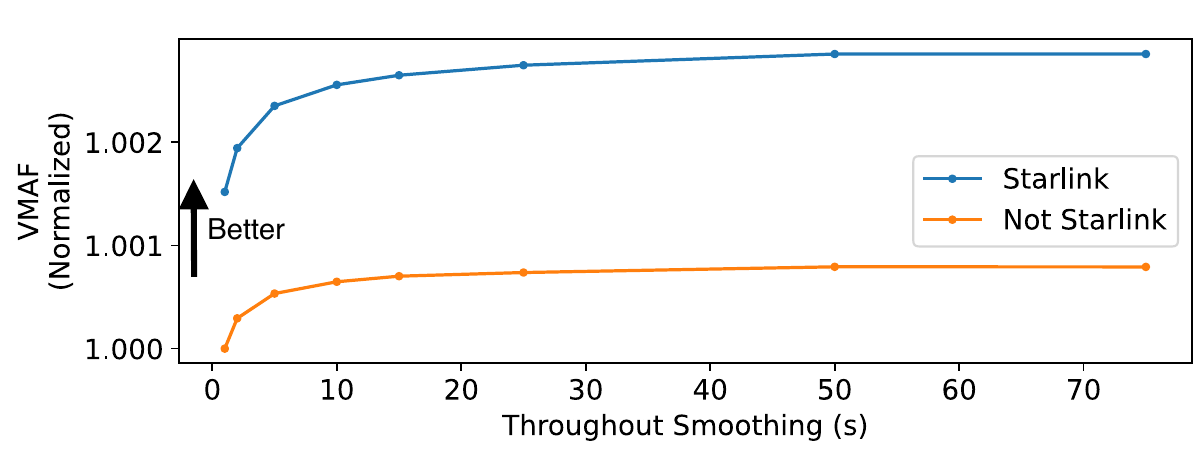}}}
        \caption{\textbf{Smoothing Effect on Perceptual Video Quality}}
        \label{fig:vmaf_smooth}
    \end{subfigure}
    \caption{\textbf{Throughput Smoothing Effects}---%
    \textnormal{While throughput smoothing helps decrease network rebuffers, Starlink's rebuffer rate never catches up to non-Starlink rebuffers. We normalize by subtracting the minimum metric (e.g., rebuffers/hour) experienced for Starlink and non-Starlink networks, respectively.}}
    \label{fig:smooth_effects}
    \vspace{-10pt} 
\end{figure*}

\vspace{3pt}
\noindent
\textbf{Throughput Smoothing.}\quad
Throughput smoothing is more likely to overestimate available throughput over Starlink, leading to increased rebuffers for Starlink sessions. 
We analyze how smoothing throughput affects throughput estimates in Figure~\ref{fig:smoothin_overestimate}, by smoothing throughput with different window sizes and calculating what percent of the time the smoothed throughput overestimates the real throughput. 
Due to Starlink's excessive variability (Section~\ref{sub:sec:variable_throughput}), smoothed throughput overestimates throughput by up to an additional 10\% compared to non-Starlink throughputs. 
Consequently, by overestimating more often, Starlink sessions are more likely to not have the assumed available throughput, leading to a potentially dangerously-large bitrate selection that is more likely cause a rebuffer.

Starlink sessions are more likely than non-Starlink sessions to experience a rebuffer, no matter the amount of throughput smoothing used. 
In Figure~\ref{fig:smooth_rebuffer}, we plot the average number of normalized network rebuffers per hour that results from using different throughput smoothing windows in our simulated ABR strategy.
We normalize by dividing the minimum metric (i.e., rebuffers/hour) experienced for Starlink and not-Starlink networks, respectively. 
Starlink sessions under the same ABR configuration, on average, experience a roughly 35\% significant increase\footnote{A t-test reveals that the rebuffer differences are statistically significantly ($p < 0.05$) above the mean of 0. } in the number of network rebuffers per hour. 
Critically, no amount of Starlink throughput smoothing actually allows it to achieve a non-Starlink equivalent network rebuffer rate. 

To minimize rebuffers, Starlink throughput requires less smoothing (i.e., need to be more reactive to sudden changes)  to trend towards non-Starlink rebuffer rates.
Across both Starlink and non-Starlink sessions, as throughput is decreasingly smoothed, the number of network rebuffers statistically significantly\footnote{Linear regression returns a p-value of less than 0.05 for both Starlink and non-Starlink sessions.\label{linear_foot}} decreases.
However, configuring ABR with Starlink-tailored parameters comes at a slight cost.
For example, reacting quicker to changing throughputs increases bitrate switches and decreases visual perceptive quality (VMAF).
\begin{figure*}[t] 
    \centering
    \begin{subfigure}[t]{0.48\textwidth} 
        \vtop{\vskip0pt 
        \hbox{\includegraphics[width=\linewidth]{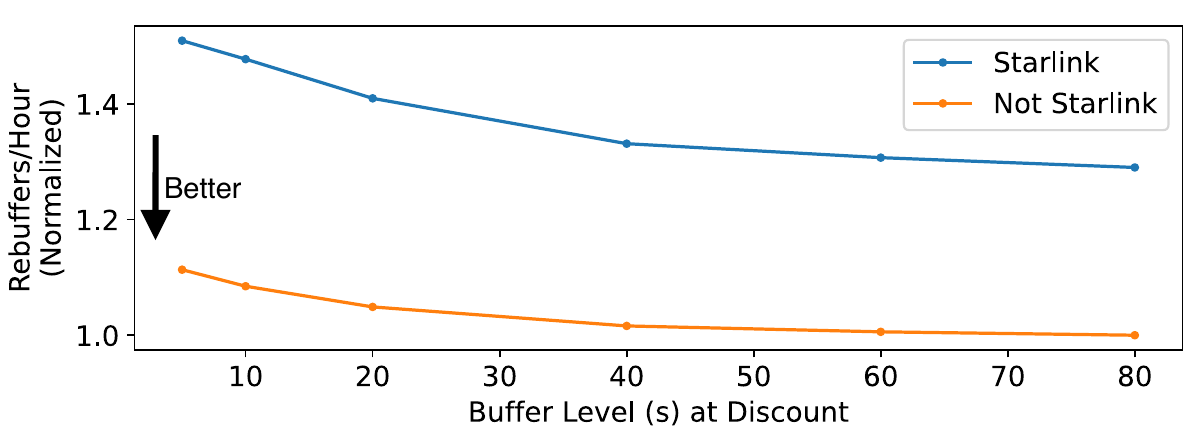}}}
        \caption{\textbf{Buffer Discount Effect on Network Rebuffers}}
        \label{fig:buffer_rebuffer}
    \end{subfigure}
    \hfill
    \begin{subfigure}[t]{0.48\textwidth} 
        \vtop{\vskip0pt 
        \hbox{\includegraphics[width=\linewidth]{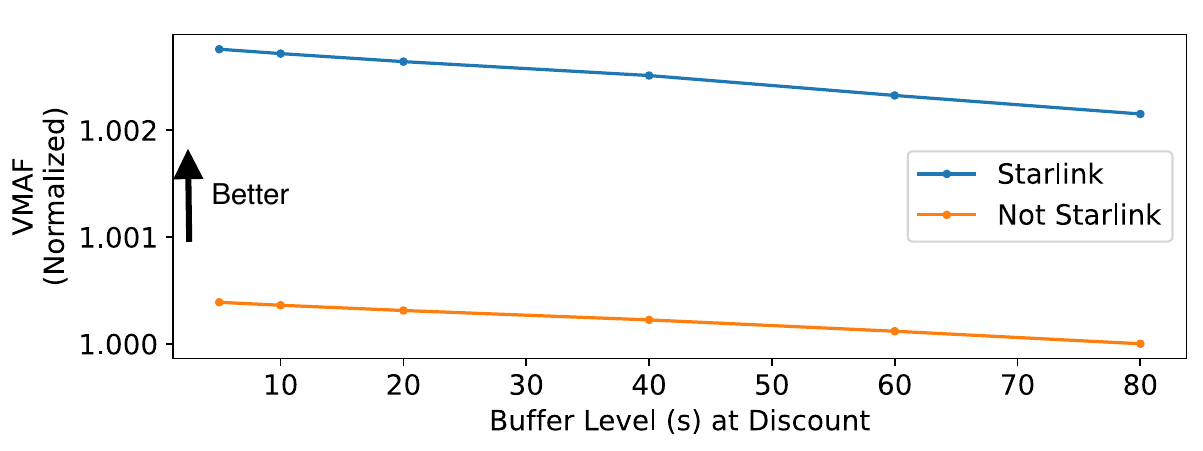}}}
        \caption{\textbf{Effect of Buffer Discount on Perceptual Video Quality}}
        \label{fig:vmaf_buffer}
    \end{subfigure}
    \caption{\textbf{Effect of Buffer Discount}---%
    \textnormal{While decreasing the buffer level at which to discount throughput prediction helps decrease rebuffers, Starlink's rebuffer rate never catches up to non-Starlink rebuffers. Decreasing buffer discounts leads to small increases in perceptual video quality (VMAF); yet, non-Starlink VMAF never catches up to Starlink VMAF.}}
    \label{fig:rebuffer_effects}
    \vspace{-10pt} 
\end{figure*}
In Figure~\ref{fig:vmaf_smooth}, we illustrate that decreasing throughput smoothing decreases\footref{linear_foot} time-weighted VMAF. 
Although the decrease in the absolute value of time-weighted VMAF is small, it can still greatly impact time weighted bitrate, as discussed in Section~\ref{sub:sec:vmaf}.


Real deployed A/B tests also reflect simulated results.
We deploy two identical ABR algorithms---with the exception of one using exponentially weighted moving average throughput smoothing of 20~seconds and another of 50~seconds---across roughly 300K randomly sampled Starlink customers streaming \netflix. 
Indeed, for sessions using the 20~second throughput smoothing, network rebuffers and time weighted VMAF decreased: rebuffers decreased by 5\%, while the bottom 10\% of time weighted VMAF decreased by 7\%. We use a Mann-Whitney test comparing the quantile distribution of the chosen metric to derive these statistically significant results.


\vspace{3pt}
\noindent
\textbf{Buffer and Throughput Discount.}\quad
A similar trend exists when configuring throughput and buffer discounts; Starlink sessions require higher discounts to trend towards non-Starlink levels of rebuffers.
For example, Starlink sessions require larger buffer levels in our simulated ABR, as depicted in Figure~\ref{fig:rebuffer_effects}, to achieve rebuffer rates closer to non-Starlink sessions. 
The larger buffer levels are likely required to accommodate Starlink sessions three times longer recovery period from throughput dips (Section~\ref{sub:sec:variable_throughput}) and 
 six times more likely to experience longer outages (Section~\ref{sub:sub:sec:net_rebuffers}).

\begin{figure*}[t] 
    \vspace{-10pt} 
    \centering
    \begin{subfigure}[t]{0.48\textwidth} 
        \vtop{\vskip0pt 
        \hbox{\includegraphics[width=\linewidth]{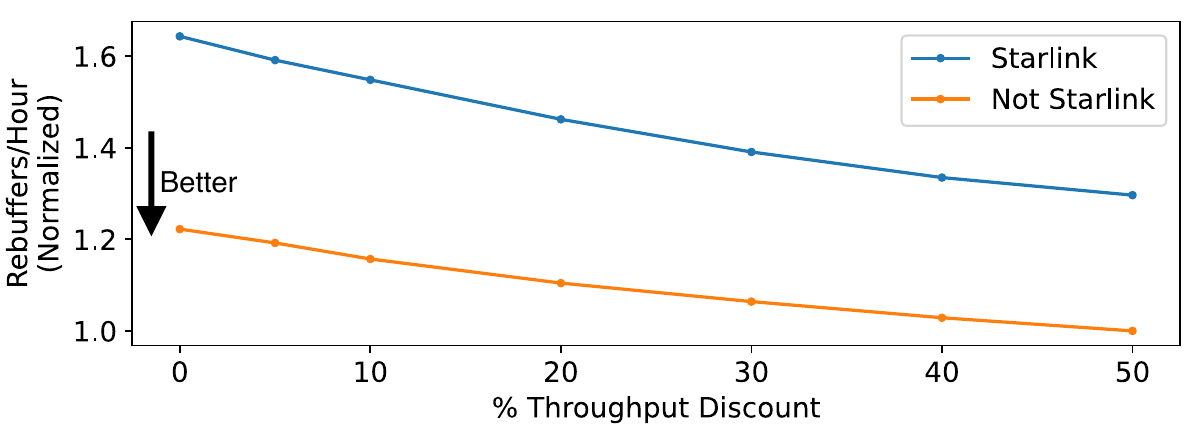}}}
        \caption{\textbf{Throughput Discount Effect on Network Rebuffers}}
        \label{fig:throughput_rebuffer}
    \end{subfigure}
    \hfill
    \begin{subfigure}[t]{0.48\textwidth} 
        \vtop{\vskip0pt 
        \hbox{\includegraphics[width=\linewidth]{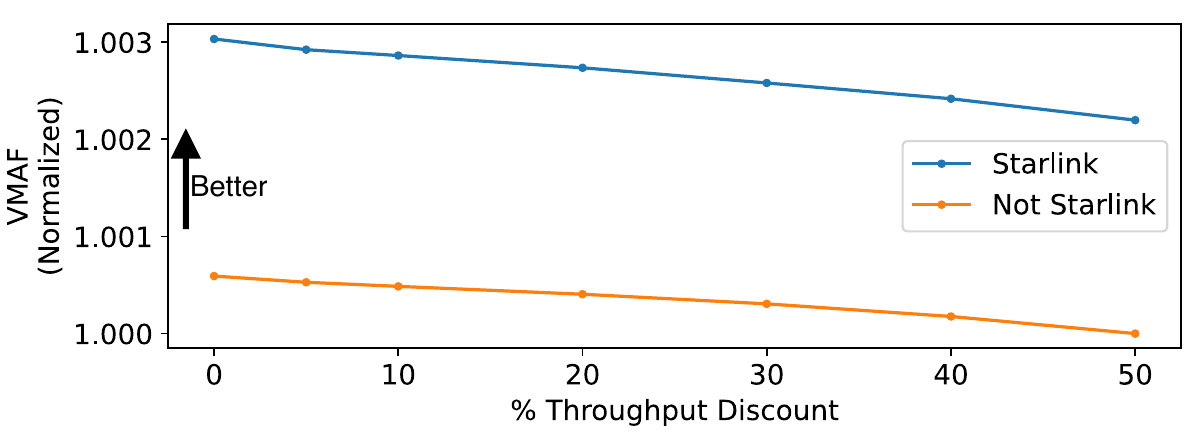}}}
        \caption{\textbf{Throughput Discount Effect on Perceptual Video Quality}}
        \label{fig:vmaf_throughout}
    \end{subfigure}
    \caption{\textbf{Throughput Discount Effects}---%
    \textnormal{While decreasing throughput discounts helps decrease rebuffers, Starlink's rebuffer rate never catches up to non-Starlink rebuffers. Further, decreasing throughput discounts leads to small increases in bad perceptual video quality (VMAF); yet, critically, non-Starlink VMAF never catches up to Starlink VMAF.}}
    \label{fig:throughput_effects}
    \vspace{-10pt} 
\end{figure*}

Meanwhile, Starlink sessions need greater throughput discounts to achieve a rebuffer rate closer to non-Starlink sessions.
Figure~\ref{fig:throughput_rebuffer}  shows that applying a 50\% Starlink throughput discount achieves roughly the same rebuffer rate as applying no throughput discount to non-Starlink sessions in our simulated ABR. 
The greater throughput discounts help mitigate the fact that 80\% of Starlink's observed throughput increase/decrease with greater magnitude than in non-Starlink networks (Section~\ref{sub:sec:variable_throughput}).
However, due to Starlink's naturally lower throughput (Section~\ref{sub:sec:throughput_less}), Figure~\ref{fig:vmaf_buffer} shows that the greater throughput discount leads to the inevitable decrease in VMAF for Starlink sessions relative to non-Starlink sessions. 




\vspace{3pt}
\noindent
\textbf{Future of ABR.}\quad
Our results show that variable throughput greatly impacts the efficacy of the three commonly deployed adaptive bitrate design principles: throughput smoothing, throughput discount, and buffer discount. 
Using these design principles, mitigating rebuffers in variable networks comes at a greater expense of other quality of experience metrics compared to less-variable networks.
Fundamentally, throughput smoothing, throughput and buffer discount all rely on point estimation that do not explicitly capture or handle high variance in throughput. 

To increase the effectiveness of ABR algorithm design, throughput variation needs to be taken out as the first-class factor and handled more deliberately. 
Throughput variation should be measured, predicted, and used directly as part of the bitrate selection. 
For example, tracking throughput variance can inform in real-time how to parameterize existing ABR strategies (i.e., how much throughput smoothing to apply and when, how much to rely on buffer levels). For ML-based ABR algorithms, this also means throughput variations need to be properly represented in the training dataset. 
\section{Limitations}
\label{sec:limitations}

Our understanding of Starlink's growth in delivering video, the quality of experience of its users, and the impact of congestion control and adaptive bitrate streaming on video streaming over Starlink are inherently biased by our vantage point: \netflix. 

This bias manifests in several ways. For instance,  trends in Starlink's growth over time (Section~\ref{sec:leo_rise}) are predominantly based on \netflix users and may not accurately represent Starlink's overall user base.
While we conclude that the number of users streaming \netflix over Starlink is larger than over other LEO networks, it is possible that users on other LEO networks are significantly less likely to stream \netflix and more likely to stream over other video platforms. 
Furthermore, our analysis of Starlink's delivery of high-quality video (Section~\ref{sec:qoe}) relies on \netflix's streaming algorithms and congestion controller and might not hold for other streaming services. Similarly, the quality of 
experience we measured for Starlink users may not translate to other types of applications, as our findings are primarily geared toward video streaming. For example, the video  buffer helps hide many of transient network disruptions typical of LEO access.

To gain a deeper understanding of LEO's growth and delivery of high quality video, we encourage the research community to conduct measurements from additional vantage points.  
Nevertheless, given that our data is collected from one of the largest streaming services, \netflix, our vantage point still accurately captures the experience of a significantly large user population.

\section{Related Work}
\label{sec:related}
Decades of research have ensured that on-demand video streaming provides a high quality of experience in the face of variable network characteristics~\cite{salehi1996supporting,jiang2012improving,spang2023sammy,yan2020learning}. 
LEO provides some of the most extreme variability in latency and throughput~\cite{izhikevich2024democratizing,mohan2023multifaceted}.
Yet,
understanding operational LEO networks and their interaction with the Internet, is still in its early infancy; LEO began serving consumers only in early 2022~\cite{reddit_star_preorder} and LEO still holds a high barrier to measurement~\cite{izhikevich2024democratizing}. 
We obtain the first global understanding of video streaming over LEO by capitalizing on our unique vantage point at \netflix, which allows us to measure over two orders of magnitude more traffic and users than the largest LEO satellite studies to date~\cite{mohan2023multifaceted,izhikevich2024democratizing}.

To adapt to variable network conditions, video streaming relies on adaptive bitrate (ABR) to select audio and video bitrates for each video chunk that maximize quality of experience (e.g., maximize bitrate, minimize rebuffers, minimize playback latency, etc.)~\cite{liu2011rate}. 
For example, some ABR algorithms use video buffer level to select bitrates that are least likely to drain the buffer~\cite{huang2014buffer, spiteri2020bola}.
Other ABRs rely on throughput measurements to inform bitrate selection~\cite{akhtar2018oboe,jiang2012improving, li2014probe}. ABR can also use a combination of both buffer levels and throughput~\cite{yin2015control, tian2012towards}. 

It is generally understood that ABR algorithms perform better when they more accurately predict the throughput~\cite{sun2016cs2p,mao2017neural} or the chunk transmission time~\cite{yan2020learning}. However, LEO throughput is known to be very unstable, which can make it hard to predict~\cite{garcia2023multi}. To the best of our knowledge, only one LEO-optimized ABR exists: Zhao~\cite{zhao2024low} designed an ABR algorithm specific for Starlink, but it suffers from increased bitrate switches and an undesirable rebuffer-quality tradeoff, similar to \netflix's non-LEO specific ABR (Section~\ref{sec:abr_design}).

While ABR algorithms address network variability by adjusting video chunk bitrates, congestion control algorithms complement this approach by focusing on packet-level adjustments. Congestion control mechanisms fine-tune the sending rate of individual packets within each chunk to maximize throughput while preventing network congestion. Decades of research have focused on adapting congestion control to various network types, including satellite networks~\cite{akyildiz2001tcp,akyildiz2002tcp, henderson1999transport,balakrishnan1997comparison,page2023distributed}. Most of this research was theoretical, conducted before large-scale LEO networks were commercially available. 

Recent advancements have renewed focus on congestion control in LEO networks. Page et al.~\cite{page2023distributed} discuss the theoretical implications of periodic re-configuration in satellite networks, which can lead to variable TCP flow counts and potentially unfair bandwidth allocation. In contrast, our empirical research shows how loss-based congestion control under-utilizes LEO links, offering a practical perspective that complements these theoretical insights. Cao et al.~\cite{cao2023satcp} introduce SaTCP, a cross-layer solution that improves TCP performance in LEO networks by predicting satellite handovers and adapting congestion control accordingly. However, such LEO-specific TCP replacements introduce significant operational complexity in environments like \netflix, where hundreds of millions of users rely on a myriad of access networks. Simulation studies of Starlink paths~\cite{hypatia,barbosa2023comparative} suggest that loss-based algorithms can achieve high throughput even with high delays, contradicting our empirical observations in Section~\ref{sec:tcp}. Garcia et al.~\cite{garcia2023multi} also report significant throughput fluctuations and severe link under-utilization in Starlink, findings that align with our results in Section~\ref{sec:qoe} and~\ref{sec:tcp}.



Researchers have significantly advanced our understanding of operational LEO networks in recent years by developing flexible open-source platforms and probing techniques that measure network characteristics across hundreds to thousands of users~\cite{zhao2024lens,ripe_atlas,izhikevich2024democratizing}. These efforts have improved our understanding of LEO's unique network characteristics and their impact on web browsing~\cite{kassem2022browser}, routing~\cite{bhosale2023characterization,wang2023reliability}, and security~\cite{giuliari2021icarus}. Additionally, small-scale studies with a few of users have begun to explore how different video streaming platforms interact with Starlink~\cite{mohan2023multifaceted,zhao2023realtime,zhao2024low}. Building on this work, our global perspective of over one million video streaming households across two years allows us to deeply understand the similarities and differences in users' quality of experience across different continents and over time.

\section{Future Work}

Looking forward, as we adapt video streaming—and the Internet more broadly—to the rise of LEO networks, it is crucial to recognize the unique challenges posed by the highly dynamic nature of networks like Starlink. Starlink's infrastructure and network are in constant flux, with ongoing enhancements aimed at improving routing, latency, bandwidth, etc.~\cite{nathan_latency_report}.
Furthermore, quality of experience dramatically varies depending upon where one is located within the Starlink network. 
Thus, as researchers and engineers adapt the Internet (e.g., congestion control, streaming algorithms) to accommodate LEO, building robust solutions will require that solutions do not overfit to how Starlink works during a particular moment in time or at a specific place in the world. 
Rather, solutions must treat Starlink as \emph{an example} of a consistently variable network that comes with constantly changing throughput, latency, packet loss, and availability, depending upon the state of the network and the relative location of the user. 

Beyond creating solutions that are robust to LEO's changing performance and globally-distributed clients, we should also strive to engineer solutions that are generalizable and adaptable to all types of networks.
Currently, many research efforts focus on optimizing for LEO-specific conditions. However, such narrowly focused solutions may not be practical in real-world applications where users access services via a mix of network types (terrestrial, LEO, GEO, mobile), and tailoring solutions to each one adds overwhelming complexity. 
As we develop new congestion control and adaptive bitrate streaming algorithms, we should build general solutions that consider non-congestive losses and treat throughput variation as a first order metric, rather than merely optimizing for performance over LEO networks. As LEO networks grow in popularity and continue to bridge the connectivity gap for both well-served and under-served populations, the development of generalized, adaptable solutions will become increasingly vital. 
These solutions will not only enhance Internet accessibility but also ensure that all users enjoy high-quality video streaming experiences, regardless of their network type.

\section{Conclusion}

In this work, we conducted the first global analysis of video streaming on demand over LEO.
We analyzed over one million Starlink users across 85~countries for over two years from the perspective of a large video streaming service, \netflix. 
We found that Starlink delivers video of similar quality compared to non LEO networks in well-served areas (e.g., US, Canada), and often better quality compared to non-LEO networks in historically under-served areas (e.g., small islands and Africa).
Nevertheless, video streaming over Starlink experiences a marginal increase in bitrate switches and rebuffers, which disproportionately affect underserved areas, and are not easily fixed by simply modifying existing congestion control and ABR algorithms. 
We hope that understanding the past and present of video streaming over Starlink helps researchers and video streaming providers adapt to a future where millions more users stream video over LEO\@.



{
\bibliographystyle{ACM-Reference-Format}
\bibliography{reference}}

\end{document}